\RequirePackage{lineno}
\documentclass[12pt]{article}
\usepackage{graphicx,color,epstopdf,grffile,upgreek}
\usepackage[bookmarks,pdfhighlight=/O,pdfstartview=FitH]{hyperref}
\usepackage{amsfonts,mathrsfs,amssymb,amsmath,amsopn,amsthm,bm,bbm,slashed,qmech}

\title{Novel physics opportunities at the HESR-Collider with PANDA at FAIR}
\author{Leonid Frankfurt$^1$, Mark Strikman$^2$, Alexei Larionov$^{3,4,5}$,\\ 
Andreas Lehrach$^{3,6}$, Rudolf Maier$^{3,6}$, Hendrik van Hees$^{7}$,\\ 
Christian Spieles$^7$, Volodymyr Vovchenko$^{5,7}$, Horst Stoecker$^{5,7,8}$
\\ $^1${\small \it Sackler School of Exact Sciences, Tel Aviv University, Tel Aviv, Israel}
\\ $^2${\small \it Pennsylvania State University, University Park, PA, USA}
\\ $^3${\small \it Institut f\"ur Kernphysik, Forschungszentrum J\"ulich, D-52425 J\"ulich, Germany}
\\ $^4${\small \it National Research Center ``Kurchatov Institute'', 123182 Moscow, Russia}
\\ $^5${\small \it Frankfurt Institute for Advanced Studies, Giersch Science Center,}
\\{\small \it D-60438 Frankfurt am Main, Germany}
\\ $^6${\small \it JARA-FAME (Forces and Matter Experiments), Forschungszentrum J\"ulich }
\\{\small \it and RWTH Aachen University, Germany}
\\ $^7${\small \it Institut f\"ur Theoretische Physik,
Goethe Universit\"at Frankfurt,}
\\{\small \it D-60438 Frankfurt am Main, Germany}
\\ $^8${\small \it GSI Helmholtzzentrum f\"ur Schwerionenforschung GmbH,}
\\{\small \it D-64291 Darmstadt, Germany}
}

\date{\today}

\begin{document}
\maketitle

\begin{abstract}
Exciting new scientific opportunities are presented for the PANDA detector at the High Energy Storage Ring in the redefined $\bar{\text{p}} \text{p}(A)$ collider mode, HESR-C, at the Facility for Antiproton and Ion Research (FAIR) in Europe. 
The high luminosity, $L \sim 10^{31}$ cm$^{-2}$ s$^{-1}$, and a wide range of intermediate and high energies, $\sqrt{s_{\text{NN}}}$ up to 30 GeV for $\bar{\text{p}} \text{p}(A)$ collisions will allow to explore a wide range of exciting topics in QCD, including the study of the production of excited open charm and bottom states, nuclear  bound states containing heavy (anti)quarks, the interplay of hard and soft physics in the dilepton production, and the exploration of the regime where gluons -- but not quarks -- experience strong interaction.
\end{abstract}

\section{Introduction}

The experimental discovery of charmonium
\cite{Augustin:1974xw,Aubert:1974js} and bottomonium \cite{Herb:1977ek}
in $\e^+\e^-$ and p$A$ collisions suggests that hadrons containing heavy
quarks can be investigated in 
hadronic  processes, where 
 dense,
strongly interacting medium could be formed. It can be particularly useful to
study the annihilation of antiprotons on free protons and baryons bound
in nuclei in $\bar{\text{p}} \text{p}(A)$ collisions, in both collider and fixed-target
experiments at collision energies of $\sqrt{s}=2$-$200 \, \GeV$.

A unique opportunity to do this in the near future is provided by the
Facility for Antiproton and Ion Research~(FAIR), with the PANDA detector
at the high-energy storage ring~(HESR). This concerns both, the
presently developed HESR fixed target mode at $\sqrt{s}<6 \, \GeV$, and
a future collider mode at $\sqrt{s}< 32\, \GeV$, with PANDA as
midrapidity detector.  The collider mode would need just an
injection-beam transfer line from the SIS 18 directly into HESR, as
discussed in Ref.~\cite{Stocker:2015cva}. Additionally, asymmetric HESR
collider schemes with somewhat lower center-of-mass energies have been
discussed in detail
\cite{Barone:2005pu,pax:2006.asy2,Lehrbach:2007.asy3}.  It may also be
feasible to study $\bar{\text{p}}A$ collisions for $\sqrt{s_{NN}}$ of up
to 19\,GeV with interesting physics opportunities \cite{Mishustin:1993qg,Larionov:2008wy}.

Luminosities of up to $5\cdot 10^{31} \, \text{cm}^{-2}\text{s}^{-1}$
can be reached at $\sqrt{s} \simeq 30 \,\GeV$ in the symmetric
$\bar{\text{p}} \text{p}$ collider mode at the
HESR~\cite{Bradamante:2005wk,Lehrach:2005ji}.  The collision scheme of
twelve proton bunches colliding with the same amount of antiproton
bunches has to be adapted to the HESR. This modification of the HESR
requires a second proton injection, the Recuperated Experimental Storage Ring (RESR), the 8\,GeV electron cooler and a modification of the
PANDA interaction region.

Besides the deceleration of rare-isotope beams, the RESR storage ring
also accumulates high-intensity antiprotons, via the longitudinal
momentum stacking with a stochastic cooling system \cite{Beller:2006sx}.
This is achieved by injecting and pre-cooling the produced antiprotons
at 3\,GeV in the Collector Ring (CR) storage
ring.

The anticipated beam intensities in the HESR proton-antiproton collider
version require a full-energy electron cooler (8\,MeV) to avoid beam
emittance growth, which results in a decreased luminosity during the
cycle. The Budker Institute of Nuclear Physics (BINP) presented a
feasibility study for magnetized high-energy electron cooling. An
electron beam up to 1\,A, accelerated in dedicated accelerator columns
to energies in the range of 4.5-8\,MeV has been proposed.  For the FAIR
full version, it is planned to install the high-energy electron cooler
in one of the HESR straight sections
\cite{Parkhomchuk:2004jc,Reistad:2006vm,Kamerdzhiev:2014yza}.

In the fixed-target mode, at $E_{\text{kin}}=4$-$10 \, \GeV$, it will be
possible to perform complementary measurements of the cross section of
charmonium interaction with nuclear matter with the PANDA detector.

A conservative estimate of the $\text{p} \bar{\text{p}}$
luminosities which can be reached at  the startup phase
without RESR is 
 $4 \cdot 10^{30} \, \text{cm}^{-2} \text{s}^{-1}$. We will use it below in our estimates assuming a one year run ($10^7 \text{sec}$).
 Energies $\sqrt{s}$ up to
$30 \,\GeV$ could be reached and  it may also be feasible to study $\bar{\text{p}}$ -
heavy-nucleus collisions for $\sqrt{s_{NN}} $ of up to 19 GeV.  In the
present work we outline how the collider at the HESR machine will extend
the scope of the PANDA project, with a focus on a few highlights.  In
Sections 2 and 3 we discuss the potential of the
$\bar{\text{p}} \text{p}$ collider to provide new information in the
field of heavy quark physics, with some attention devoted to the
possible discovery of new states. Some other related opportunities are
outlined in Sec.\ 4.  Section 5 presents a number of additional physics
topics that could be explored in both $\bar{\text{p}} \text{p}$ and
$\bar{\text{p}} A$ modes.  These include the production of nuclear
fragments containing $\bar{\text{c}}$ and/or c quarks, the
$\text{c} \bar{\text{c}} $ pair production, color fluctuation effects,
probing the pure glue matter, and the production of low-mass dileptons.
Concluding remarks in Sec.~6 close the article.

\section{Study of bound states containing heavy quarks}

A number of new states containing heavy quarks have been discovered
recently.  These can be interpreted as pentaquark and tetraquark states
containing $\text{c} \bar{\text{c}}$ pairs.  Some of the states are
observed in decays of mesons and baryons containing b-quarks, others in the final states of
$\e^+\e^-$ annihilation; for a review see \cite{Karliner:2015afa}. 

It is widely expected that these discoveries represent just the start of
the exploration of rich new families of states containing heavy quarks.
Understanding the dynamics responsible for the existence of these states
would help to clarify many unresolved issues in the spectroscopy of
light hadrons.

A unique feature of an intermediate-energy $\bar{\text{p}} \text{p}$
collider is that it makes
possible to study the production of
$Q\bar Q$ ($Q = \text{c},\text{b}$) pairs and the formation of various
hadrons containing heavy quarks rather close to the threshold.  The
$\text{b} \bar{\text{b}}$ pairs are produced mostly in the process of
annihilation of valence quarks and antiquarks, i.e.
$q\bar q \to Q\bar Q$.  The production of $\text{c}\bar{\text{c}}$ pairs in
the antiproton fragmentation region also corresponds to this mechanism.

The invariant masses of the produced $Q\bar Q$ pairs are much closer to
the threshold in the discussed energy range than at the LHC energies. It is
natural to expect that the large probability to produce final states
with small $Q\bar Q$ invariant masses should lead to a higher relative
probability to produce pentaquark and tetraquark states compared to the
one at the LHC energies.  Additionally, the small transverse momenta of
the $Q\bar Q $ pairs facilitate the pick up of light quarks as
compared to $Q$ or $\bar{Q}$ fragmentation.  In the
antiproton-fragmentation region another $Q\bar Q $ production
enhancement mechanism, specific for antiproton interactions, is
possible: the production of $Q\bar{Q}$ pairs with large
$x \sim 0.2$-$0.4$ in the annihilation of $q \bar{q}$, which could merge
with a spectator antiquark of the antiproton carrying $x\sim 0.2$.

Another effect which can help to observe new states in medium-energy
$\bar{\text{p}}\text{p}$ collisions is the relatively low spatial
density of the system produced at moderate energies. This should
suppress final-state interactions, which could possibly hinder the formation of
weakly bound clusters of large size. An additional advantage is a relatively small
bulk hadron production which reduces the
combinatorial background significantly as compared to the LHC.

\subsection{The heavy quark production rates} 

The need for $\text{t} \bar{\text{t}}$-production cross
sections has stimulated the development of
new computational techniques for heavy-quark production in hadron-hadron
collisions~(see, e.g., Refs.~\cite{Cacciari:2005uk,Beneke:2002ph}), in
particular those which include the effects of threshold resummation.

These calculations, which are currently being validated by comparison to
data at high collision energies, predict the following cross section for
the $\text{b} \bar{\text{b}}$ pair production in $\bar p p$ collisions \cite{CacciariVogt}:
\begin{equation}
  \sigma_{\text{b}\bar{\text{b}}} (\sqrt{s} = 30 \,\GeV)= 1.8 \cdot
  10^{-2} \, \upmu\text{b}.
  \label{1}
\end{equation}
This calculation has a relative uncertainty of about 30\%, and the
predicted cross section value is about seven times higher than the
corresponding cross section for pp scattering 
because of the contribution of the valence-quark valence-antiquark
annihilation present in $\bar{\text{p}} \text{p}$.  The
$\text{b}\bar{\text{b}}$ cross section per nucleon for the $15~\text{GeV} \times 6~\text{GeV}$
kinematics is a factor of 100 smaller. 
Charm production is dominated by gluon
annihilation, $gg \to Q\bar Q$. This fact implies that the corresponding
cross sections are close in pp and $\text{p}\bar{\text{p}}$ collisions,
with the exception of the fragmentation regions.  The experimental data
in this case are rather consistent between pp and
$\bar{\text{p}} \text{p}$ and correspond to 
\begin{equation}
  \sigma_{\text{c}\bar{\text{c}}}= 30\,\upmu\text{b}.
  \label{2}
\end{equation}
For the $15~\text{GeV} \times 6~\text{GeV}$ scenario the cross section per nucleon drops by a
factor of 3.

\subsection{Rate estimates}

The cross sections in Eqs. (\ref{1}) and (\ref{2}) correspond to
significant event rates for one year ($10^7 \, \text{s}$) of running at
a luminosity of $4\cdot 10^{30}\,\text{cm}^{-2}\text{s}^{-1}$.  We find
\begin{equation}
  N_{\text{b} \bar{\text{b}}} = 10^6, \quad N_{\text{c} \bar{\text{c}}} = 10^9.
\end{equation}
These numbers can be easily rescaled for a run at a different
luminosity, if required.

At these energies a rearrangement of the light-quark fractions occurs in
the final states without significant suppression since heavy quarks are
dominantly produced at mid rapidity, and the interaction between a b
quark and a light valence quarks produces slow light quarks. Thus the
overlapping integral between such a $\text{b}\bar q$ pair and the
corresponding hadron wave function should be large. Most likely mesons
are produced in excited states. The probability for the formation of
b$\bar q$ mesons with a given flavor is about $30\%$ of the total cross
section because of the competition between different flavors. In the
discussed processes gluon radiation is a small correction because of the
restricted phase space and the large b-quark mass.

At the collision energies considered here, the combinatorial background is
significantly smaller compared to LHC energies, which makes the
observation of the excited states containing heavy quarks easier.  In
particular, the spectator quarks and antiquarks will have quite small
velocities relative to the $\text{b}$ quark.
Indeed, for $\sqrt{s} = 30 \,\GeV$ a typical $x$ value for
the b-quark is $m_{\text{b}}/\sqrt{s} \gtrsim 0.2$, which is close to the
$x$ values of the valence quarks.  This fact suggests the possibility of
formation of excited states in both the meson and the baryon channels,
as well as an enhanced probability of the production of the
excited $\text{b} \bar{\text{b}} q \bar q$ tetraquark or $(\text{b}\bar q)-(\bar{\text{b}} q)$
mesonic molecular states.

\subsection{Hidden beauty resonance production}

In contrast to charm production, the cross section for hidden beauty
resonance production $\text{p} \bar{\text{p}} \to \chi_{\text{b}}$ may
be too low for the process to be observed at the HESR. 
Within the standard quarkonium models, 
where a $Q\bar Q$ pair annihilates into two gluons which subsequently fragment into light quarks, 
one can estimate that this cross section drops with $M_Q$ as $ R_{\text{b/c}}=\Gamma(\chi_{\text{b}} \to \text{p} \bar{\text{p}})/\Gamma_{\text{tot}}(\chi_{\text{b}}) \propto \alpha_s^8/M^8_Q. $ 
This is because this cross section is proportional to the partial width of the decay $\text{p} \bar{\text{p}} \to \chi_{\text{b}}$, which drops with an increase of $M_Q$.
The ratio of cross sections of beauty production
through a $\chi_{\text{b}}$ intermediate state to that for charm is
$\approx [\alpha_s(M_Q)/\alpha_s(M_c)]^8/ [M_c/M_Q]^{10}$ with an
additional factor of $M_Q^{-2}$ stemming from the expression for the
resonance cross section. The suppression is due to the necessity of a
light-quark rearrangement in the wave function of the proton to obtain
decent overlapping with $\chi_{\text{b}}$ states. In the
non-relativistic approximation the wave functions of $\chi_{\text{b}}$
states vanish at zero inter-quark distance. Thus the overall suppression
for the total cross section of $\chi_{\text{b}}$ production as compared to that of
$\chi_{\text{c}}$ production is approximately $ 10^{-7}$.

\section{Potential for discovery of new states}

Investigating $\text{p} \bar{\text{p}}$ collisions at moderate energies
carries specific advantages for searches of new states as the
$\text{b} \bar{\text{b}}$-production rate is relatively high, while the
overall multiplicity, which determines the background level, is rather
modest.  Also, an equal number of states containing quarks and
antiquarks is produced enabling cross checks of observations using
conjugated channels.  In the following we will discuss resonances
containing heavy quarks with the understanding that everything said
equally applies to the resonances containing heavy antiquarks. Although
the rates in many cases are rather modest, we nevertheless include the
discussion of these channels in view of the possibility to have a higher
energy collider, as discussed in the final remarks (Sec.\ 6).

\subsection{$\text{b}qq$-baryons and $\text{b}\bar q$ mesons}

The current knowledge of the spectrum of the excited states containing b- or $\bar{\text{b}}$-quarks is very limited.
According to PDG \cite{Tanabashi2018}, in the $q \bar b$ sector there are two states $B_J(5970)^+$ and $B_J(5970)^0$
with unknown quantum numbers which could be excited states of the $B^+$ and $B^0$, respectively. 
In the $\text{b}qq$-sector there are two baryons $\Lambda_b(5912)^0$ and $\Lambda_b(5920)^0$ which can be regarded
as orbitally excited states of $\Lambda_b^0$ and one excited $\Sigma_b^*$ state. This is much less in comparison
with the $\text{c}qq$ sector where five excited $\Lambda_c^+$ states and two excited $\Sigma_c$ states (all rated with ***)
are observed.

So there are plenty of opportunities here.  One noteworthy issue is the
comparison of the accuracy with which the heavy-quark limit works for
hadrons containing $\text{b}$ quarks vs. those containing $\text{c}$
quarks.

\subsection{Excited states containing $\text{b} \bar{\text{b}}$}

As argued above it is very difficult to produce bound states containing
$\text{b} \bar{\text{b}}$ in the resonance process of
$\text{p} \bar{\text{p}} $ annihilation.  Nevertheless, many of these
states, as well as other states like analogs of X, Y, Z charmonium
states, could be produced in inelastic $\text{p} \bar{\text{p}}$
interactions.  This is because the invariant mass of the produced
$\text{b} \bar{\text{b}} $ system is rather close to the threshold, and
because the $\text{b} \bar{\text{b}}$ pair is produced in association
with several valence quarks and valence antiquarks which have rather low momenta
relative to the $\text{b} \bar{\text{b}}$  pair.

\subsection{Baryons and mesons containing two heavy quarks}

Since there are three valence antiquarks colliding with three valence
quarks in $\bar{\text{p}} \text{p}$, one can produce two pairs of
heavy quarks in a double quark-antiquark collision.  This entails a
possibility for producing the following baryons and mesons containing
two heavy quarks:

\textit{(i) $\text{cc}q$ baryons}.

At the collision energies discussed, the contribution of the
leading-twist mechanism of $2g\to Q\bar Q Q\bar Q$ for the double
heavy-quark production should be quite small as it requires very large
$x$ of the colliding partons (the situation might be less pronounced for
the case of the double $\text{c} \bar{\text{c}}$ production than for
$\text{b} \bar{\text{b}}$ ).  Therefore, the only effective mechanism
left is the production of two pairs of heavy quarks in two hard
parton-parton collisions (see Fig.~1).
\begin{figure}[t]  
\centering
 \includegraphics[width=.6\textwidth]{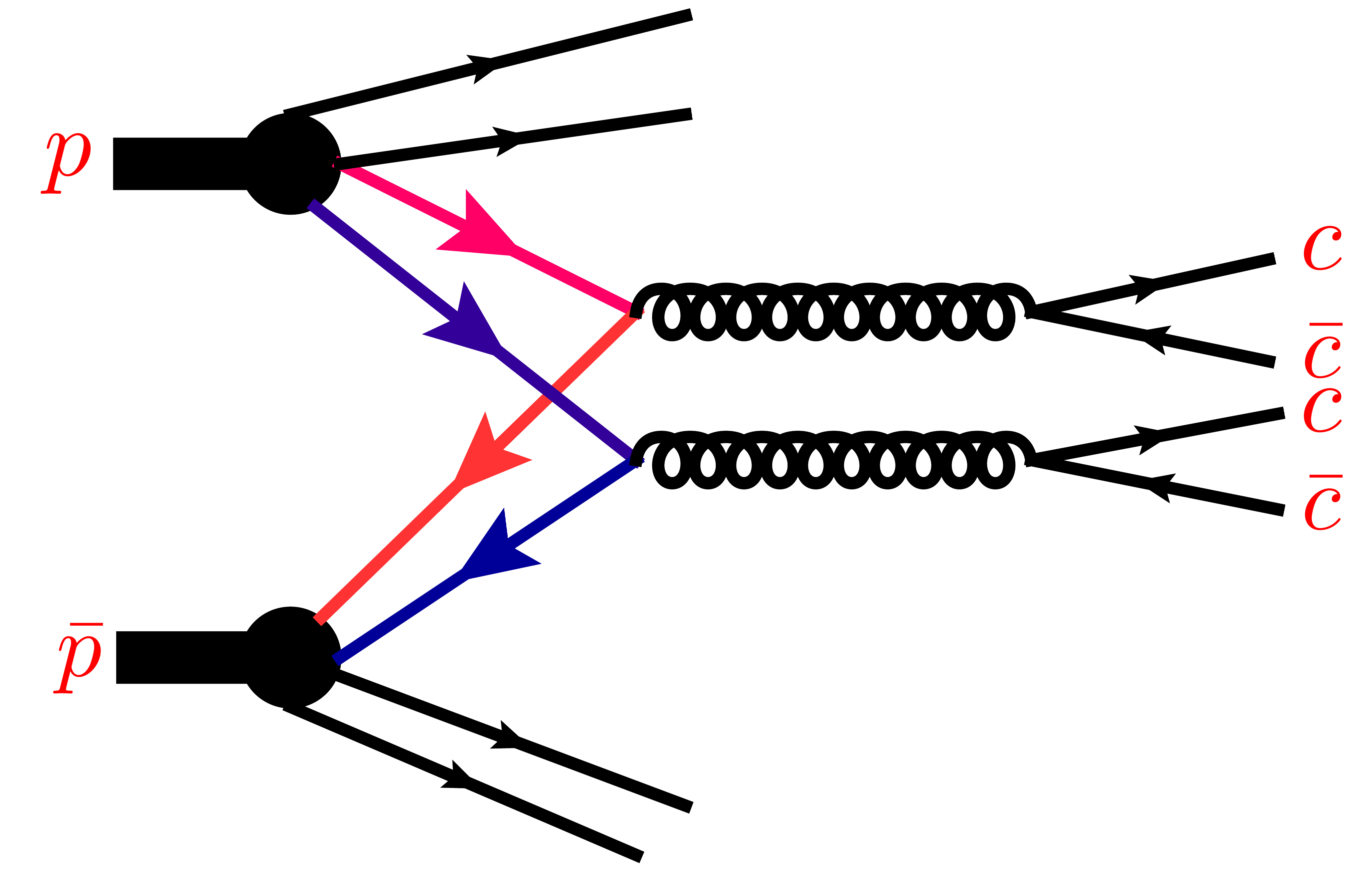}
 \caption{Double parton interaction mechanism for the production of two
   pairs of heavy quarks.}
\label{regge}
\end{figure}  

For the case of double $\text{c} \bar{\text{c}}$ pair production, one
can make estimates by considering the suppression factor for the
production of the second $\text{c} \bar{\text{c}}$ pair relative to a
single $\text{c} \bar{\text{c}}$ pair.  This factor can be roughly
estimated using the high-energy experimental studies of double-parton
collisions at the Tevatron collider. One finds a probability of about
$ 10^{-3}$ for the ratio of the cross section for producing two
$\text{c} \bar{\text{c}}$ pairs relative to a single pair.  (To be
conservative we took a factor of two smaller value of the parameter
which determines the probability of double collisions
($1/\sigma_{\text{eff}}$) than the one measured at Tevatron as in the
Tevatron kinematics special small-$x$ effects may enhance this factor.)
These considerations result in
\begin{equation}
  N_{\text{c} \bar{\text{c}}, \text{c} \bar{\text{c}}} = 10^6
\end{equation}
as an estimate for the yearly number of events with two pairs of
$\text{c} \bar{\text{c}}$. Since the available phase space is rather
modest, there is a significant probability that the relative velocity of
two c - quarks would be small, and therefore a $ccq$ state would be
formed.

Other interesting channels are production of an open charm-anticharm pair plus charmonium, and double charmonium.

{\it  (ii) $\text{bc}q$ baryons and $\text{b}\bar{\text{c}}$ mesons}.

The $\text{b}\bar{\text{b}} \text{c} \bar{\text{c}}$ pairs are produced pretty close to threshold and have small relative velocities. 
Hence there is a good chance that they would form a $\text{bc}q$ baryon. 
About $10^3$ events per year of running with $\text{b} \bar{\text{b}} \text{c} \bar{\text{c}}$ could be expected based on the double parton interaction mechanism.

{\it (iii) $\text{bb}q$ baryons.}

To observe the production of $\sim 10^2$
$\text{b} \bar{\text{b}}\text{b} \bar{\text{b}}$ pairs one would need a
one-year run at a much higher luminosity of
10$^{33} \,\text{cm}^{-2} \text{s}^{-1}$. The velocities of two b quarks
are expected to be close in about 1/2 of the events. So there should be
a significant chance for them to form $\text{bb}q$ baryons.
 
{\it (iv) $\text{bcc}$ baryons.}

Whether it is feasible to observe $\text{bcc}$ states requires more
detailed estimates and may depend on the structure of the three-quark
configurations in the nucleon. A naive estimate is that $10^2$ events
with $\text{bcc}\bar{\text{b}} \bar{\text{c}} \bar{\text{c}}$ would be
produced in a one-year run at a higher luminosity of
$10^{33} \,\text{cm}^{-2} \text{s}^{-1}$.

Note also that even though a $\text{b} \bar{\text{c}}$ meson has been
observed (although its quantum numbers are not known), there are
potentially many other states built of these quarks suggesting a rich
spectroscopy (mirroring the spectra of $\text{c} \bar{\text{c}}$- and
$\text{b} \bar{\text{b}}$-onium states).

\subsection{Summary}

To summarize, it would be possible with PANDA at the HESR-C $\bar{p}p$ collider to discover and to study properties
of meson and baryon states containing one b quark and light (anti)
quarks, complementing the charmonium states which are planned to be
explored in the PANDA fixed target experiment. There are also good chances to
discover the double-heavy-quark baryonic and mesonic states. These new
potential observations will allow to achieve a much deeper understanding
of the bound-state dynamics in QCD.

\section{Other opportunities}

In the last decades much effort has been devoted to studying high-energy
properties of QCD in the vacuum channel - the so called perturbative
Pomeron.  Interactions in non-vacuum channels, on the other hand, have
practically not been studied. The PANDA experiment at a  collider
has a perfect kinematic coverage to study the behavior of Regge
trajectories in both the non-perturbative regime (small Mandelstam $t$)
as well as the possible onset of the perturbative regime. One advantage
of the antiproton beam is the possibility to study a wide range of
baryon and meson Regge trajectories, possibly including Regge
trajectories with charmed quarks. The latter will allow to check the
non-universality of the slopes of the Regge trajectories which were
observed already at positive $t$ values.

Other possibilities include Drell-Yan-pair measurements, which at the
lower end of the discussed energy range, 
may be extended to the limit of exclusive processes like
$\bar{\text{p}} \text{p} \to \upmu^+\upmu^- + \text{meson}$ which are
sensitive to generalized parton-distribution functions, etc.

Another direction of studies is the investigation of correlations
between valence (anti)quarks in (anti)nucleons using multi-parton interactions (MPI)
analogous to those shown in Fig.\ 1. In the discussed energy range MPI
get a significant contribution from collisions of large-$x$ partons
(double Drell-Yan, Drell-Yan + charm production, etc.), and the rate of
the MPI is inversely proportional to the square of the average distance
between the valence quarks. In particular the rates would be strongly
enhanced in the case of a large probability of (anti)quark-(anti)diquark
configurations in the (anti)nucleon.

The analysis of the production of heavy-quark pairs discussed above would 
also be of great interest for the study of the multiparton structure of
nucleons.

\subsection{$\bar{\text{p}} A$ elastic scattering and absorption}
\label{pbarA_el}

So far measurements of the antiproton-nucleus elastic scattering were
done only at LEAR for $p_{\rm lab} < 1 \,\GeV/c$. It turned out that
owing to the forward-peaked $\bar{\text{p}} \text{p}$ elastic scattering
amplitude the Glauber model describes LEAR data on the angular
differential cross sections of $\bar{\text{p}} A$ elastic scattering
surprisingly well.  This is in contrast to the $\text{p} A$ elastic
scattering where the Glauber model description starts to work only above
$p_{\rm lab} \sim 1.5$ GeV/c~\cite{Alkhazov:1978et}.

Glauber theory analysis \cite{Larionov:2016xeb} has shown that the
$\bar{\text{p}} A$ and $\text{p} A$ angular differential elastic
scattering cross sections at 
$p_{\rm lab} = 10$ GeV/c (fixed target PANDA) strongly differ
in the diffraction minima due to the different
ratios of the real-to-imaginary parts of $\bar{\text{p}} N$ and
$\text{p} N$ elastic scattering amplitudes. Experimental confirmation of
such a behavior would be a good validity test of the Glauber theory,
important in view of its broad applications for other reaction channels,
and of the input elementary amplitudes which are typically given by
Regge-type parameterizations. In particular, the $\bar{\text{p}} n$
elastic amplitude is accessible only by scattering on complex
nuclei. The determination of diffractive structures at
$\bar{\text{p}}A$-collider energies would require good transverse
momentum transfer resolution $\sim 10 \MeV/c$ and the capability to
trigger 
on the events where nucleus remains intact
(see the discussion in
Sec.~\ref{new_hq_states}).  Light nuclear targets are preferred as their
diffractive structures are broader in $p_t$, and there is a
smaller number of possible excited states.
Spin 0 targets like $^4$He are especially good for these purposes.

A related problem is the determination of the antiproton absorption
cross on nuclei (defined as the difference between total and elastic
cross sections). Experimental data on the antiproton absorption cross
section above LEAR energies are quite scarce, although such data are
needed for cosmic ray antiproton flux calculations
\cite{Moskalenko:2001ya}.

\subsection{Coherent hypernuclei production}
\label{cohHypProd}

While ordinary $\Lambda$-hypernuclei were discovered long ago, the
$\Lambda_{\text{c}}^+$- and $\Lambda_{\text{b}}^0$-hypernuclei were
predicted in mid-70s \cite{Tyapkin:1976,Dover:1977jw} but have not been observed so far. 
However, their existence is expected based on a number of models, e.g the quark-meson coupling model~\cite{Tsushima:2002ua}.

The processes $\bar{\text{p}} \text{p} \to \bar Y Y$, where
$Y=\Lambda, \Lambda_{\text{c}}^+$ or $\Lambda_{\text{b}}^0$, have the lowest thresholds
among all possible other channels of the respective $\bar{\text{s}} \text{s}$,
$\bar{\text{c}} \text{c}$ or $\bar{\text{b}} \text{b}$ production channels in $\bar{\text{p}} \text{p}$
collisions. Thus, they are preferred for $Y$-hypernuclei production
as the momentum transfer to the hyperon is relatively small.

The coherent reactions
${}^AZ(\bar{\text{p}},\bar\Lambda){}^A_\Lambda(Z-1)$ for the different
states of the hypernucleus have 
never being studied experimentally. It is expected that these reactions have
cross sections of the order of a few 10\,nb at $p_{\text{lab}} \sim 20 \,\GeV/c$ \cite{Larionov:2017hcm}.
Thus, they can serve as a powerful source of $\Lambda$-hypernuclei production at the lower end of the
$\bar{\text{p}}A$-collider energies.  Here, the amplitude
$\bar{\text{p}} \text{p} \to \bar{\Lambda} \Lambda$ should be dominated by the
$K^*$ (or $K^*$ Regge trajectory) exchange.

More challenging is the coherent 
process
${}^AZ(\bar{\text{p}},\bar\Lambda_c^-){}^A_{\Lambda_c^+}Z$
\cite{Shyam:2016uxa} where the underlying
$\bar{\text{p}} \text{p} \to \bar\Lambda_{\text{c}}^-
\Lambda_{\text{c}}^+$ amplitude is due to $D^0$ and $D^{*0}$ exchanges.
One can also think of the
${}^AZ(\bar{\text{p}},\bar\Lambda_{\text{b}}^0){}^A_{\Lambda_{\text{b}}^0}(Z-1)$ coherent
reaction.

\section{Unique opportunities for probing QCD properties at the $\bar{\text{p}} A$ collider}

\subsection{Space-time picture of the formation of hadrons containing heavy quarks}

The kinematics of heavy-quark-state production 
in collisions of p($\bar{\text{p}}$) with proton or nuclei~(neglecting Fermi motion effects)
dictates that heavy states can only be produced with momenta
\begin{equation}
p_Q >  M_Q^2/2m_Nx_q - m_Nx_q/2
\end{equation}
in the rest frame of the nucleus. Here $x_q$ is the $x$ of the quark of
the nucleus involved in the production of the $Q\bar Q$ pair. For
$x_q\le 0.5$ this corresponds to a charm momentum above $4\,\GeV/c$
which is much larger than typical momenta of the heavy system embedded
in the nucleus.

However, there is a significant probability that $\D$,
$\Lambda_{\text{c}},\ldots$ hadrons slow down due to final-state
interactions. Indeed, it is expected in QCD that the interaction
strength of a fast hadron with nucleons is determined by the area in
which the color is localized. For example, $\psi'$-$N$ interactions
should be comparable to the kaon-nucleon cross section and be much
larger than the $\text{J}/\psi$-$N$ cross section, see for example
\cite{Gerland:1998bz}. Also the cross sections of open charm (bottom)
interactions should be on the scale $\gtrsim 10 \,\text{mb}$.

The formation distance (coherence length)  can be estimated as 
\begin{equation}
l_{\text{coh}} \simeq  \gamma l_0,  
\end{equation}
where $l_0 \simeq 0.5$-$1.0 \,\fm$.  
For the discussed
energies and the case of scattering off heavy nuclei the condition
\begin{equation}
l_{\text{coh}} \leq R_A,  
\end{equation}
is satisfied for hadrons produced in a broad range of momenta including
the central and nucleus fragmentation region. So it would be possible to
explore the dependence of the formation time and interaction strength on, for
example, the orbital angular momentum of $\D^*$.

Observing these phenomena and hence exploring QCD dynamics in a new
domain could be achieved by studying the $A$-dependence of charm
production at momenta $\lesssim 10 \,\GeV/c$ (in the rest frame of the
nucleus).
 
\subsection{New heavy-quark states}
\label{new_hq_states}

The formation of heavy mesons or baryons, H, inside nuclei implies
final-state interactions which slow down these heavy hadrons, leading to the
production of hadrons at low momenta forbidden for scattering off a free
proton:
\begin{equation}
p_{\text{H}} \leq (m_{\text{H}}^2-m_N^2)/2m_N.
\end{equation}
In this kinematics the slow-down may be sufficient to allow for the
production of (anti-)charm quarks embedded in nuclear fragments. The
collider kinematics would make it easier to detect decays of such nuclei
than in fixed-target set ups as these nuclei would be produced with
high momenta (velocities comparable to those of ordinary nuclear
fragments). Thus the discussed HESR-C collider in the $\bar{\text{p}} A$ mode would
have a high discovery potential for observing various nuclear states
containing c and/or $\bar{\text{c}}$.

Higher luminosities and higher collider energies will allow search for analogous b and/or $\bar{\text{b}}$ states.

\subsection{Color fluctuations in nucleons}

At high energies hadrons are thought to be interacting with each other
in frozen configurations which have different interaction strengths --
so-called color fluctuations. One can explore these phenomena in
proton-nucleus collisions in a number of ways. Here we give as one
example the study of the interaction strength of a hadron in the case of
a configuration that contains a large-$x$ ($x \geq 0.4$) parton. One
expects that in such configurations the average interaction strength is
significantly smaller than on average: in these configurations color
screening leads to a suppression of the gluon fields and of the quark-antiquark
sea \cite{Frankfurt:1985cv}. This picture has allowed to explain
\cite{Alvioli:2014eda,Alvioli:2017wou} strong deviations of the
centrality dependence of the leading-jet production from the geometrical
picture (Glauber model of inelastic collisions) observed at the LHC in
p-Pb collisions and at RHIC in d-Au collisions.
 
Due to a fast increase of the interaction strength for small-size
configurations with increasing energy, the strength of color
fluctuations drops at higher energies. Correspondingly color-fluctuation
effects are expected to be much enhanced at the HESR-C $\bar{\text{p}} A$-collider
energies. For example, for $x \simeq 0.6$, the cross-section ratio,
$\sigma_{\text{eff}}(x)/\sigma_{\text{tot}}(NN)$, is expected to be
$\sim 0.25$, while at the LHC it is $\sim 0.6$.

To observe this effect one would need to study Drell-Yan production at
large $x$. A strong drop of hadron production in the nucleus
fragmentation region would be a strong signal for the discussed
effect. For its detailed study measurements with different nuclei would
be desirable.

\subsection{Probing pure glue matter}

One of the central questions in high-energy hadronic and nuclear
collisions is how the initially non-equilibrium system evolves towards a
state of apparent (partial) thermodynamic equilibrium at later stages of
nuclear collisions. Presently, the community favors a paradigm of an
extremely rapid ($t_{\rm eq}$ less than $0.3\,\fm/c$) thermalization and
chemical saturation of soft gluons and light quarks.

The large gluon-gluon cross sections lead to the idea
\cite{VanHove:1974wa} that the gluonic components of colliding nucleons
interact more strongly than the quark-antiquark ones. The two-step
equilibration scenario of the quark-gluon plasma (QGP) was proposed in
\cite{Raha:1990dn,Shuryak:1992wc,Alam:1994sc}.  It was assumed that the
gluon thermalization takes place at the proper time
$\tau_{\text{g}}<1~\textrm{fm}/c$ and the (anti)quarks equilibration occurs at
$\tau_{\rm th}>\tau_{\text{g}}$. The estimates of Refs.~\cite{Biro:1993qt,Elliott:1999uz,Xu:2004mz} show that
$\tau_{\rm th}$ can be of the order of $5\,\fm/c$.

\begin{figure}[t]
\centering
\includegraphics[width=0.70\textwidth]{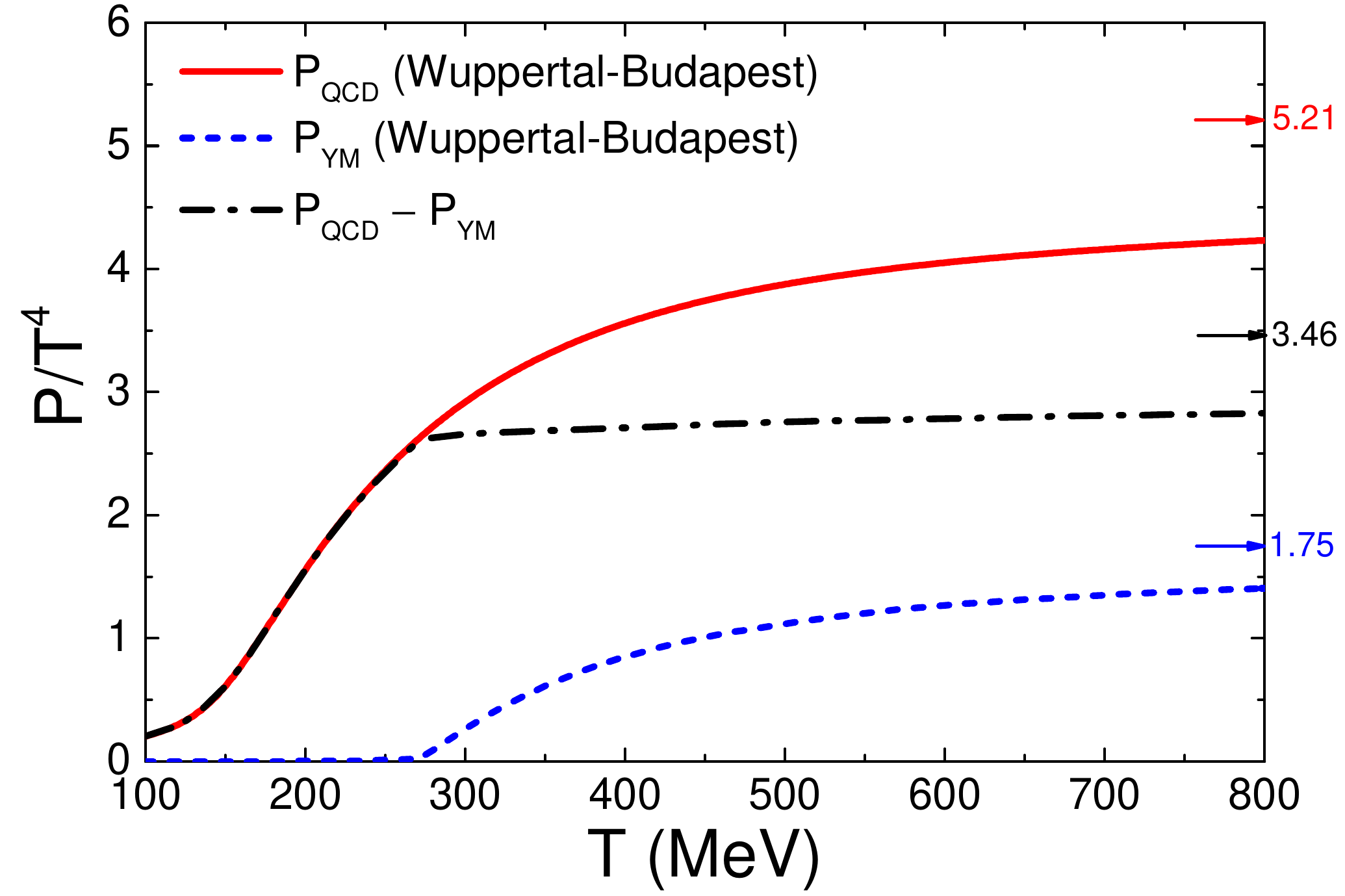}
\caption[]{Temperature dependence of the scaled pressure $p/T^4$
  obtained in lattice QCD calculations of the Wuppertal-Budapest
  collaboration for (2+1)-flavor QCD \cite{Borsanyi:2013bia} (red line)
  and for Yang-Mills matter \cite{Borsanyi:2012ve} (blue line). The
  black dash-dotted line depicts the difference between the pressure in
  full QCD and in Yang-Mills theory.  }\label{fig:latticeEoS}
\end{figure}

Recently the \emph{pure glue} scenario was proposed for the initial state at midrapidity in Pb+Pb
collisions at Relativistic Heavy Ion Collider (RHIC) and Large Hadron
Collider~(LHC) energies~\cite{Stoecker:2015zea,Stocker:2015nka}.  According to lattice-QCD
calculations~\cite{Borsanyi:2012ve}, quarkless purely gluonic matter
should undergo a first-order phase transition at a critical temperature
$T_{\text{c}}= 270 \,\MeV$. At this temperature the deconfined pure glue
matter transforms into the confined state of pure Yang-Mills theory,
namely into a glueball fluid. This is in stark contrast to full QCD
equilibrium with (2+1) flavors, where a smooth crossover transition
takes place (see Fig.~\ref{fig:latticeEoS} for a comparison of the
corresponding equations of state).

At $\sqrt{s_{NN}} \simeq 30$~GeV $\bar{\text{p}}\text{p}(A)$ collisions can create only small systems.
Baryon free matter can be expected if the $\bar{\text{p}}\text{p}$ annihilation occurs briefly in the initial stage of the collision.
An enhanced annihilation probability can be expected in $\bar{\text{p}}A$ collisions over the $\bar{\text{p}}\text{p}$ collisions.
Therefore, the properties of this baryon-free matter can potentially be studied by looking into the difference in observables between $\bar{p}p(A)$ and $pp(A)$ at the same energy.

If indeed a hot thermalized gluon fluid, initially containing no
(anti)quarks, is created in the early stage of a
$\bar{\text{p}}\text{p}$ or $\bar{\text{p}}A$ collision at mid rapidity, it will quickly cool
and expand until it reaches a mixed-phase region at $T = T_c^{\rm YM}$. 
After the initial pure gluon plasma has completely transformed into the glueball
fluid, the system will cool down further. 
These heavy glueballs produced
during the Yang-Mills hadronization process where the pure glue plasma forms a glueball fluid. 
The heavy glueballs will later evolve into lighter states,
possibly via a chain of two-body decays~\cite{Beitel:2016ghw}, and finally decay into hadronic resonances and light hadrons, which may or may not show features of chemical equilibration.

Of course, a more realistic scenario must take into account
that some quarks will be produced already before and during the Yang-Mills driven
first-order phase transition.  
This scenario can be modeled by
introducing the time-dependent effective number of (anti)quark degrees of
freedom, given by the time-dependent absolute quark fugacity
$\lambda_q$~\cite{Vovchenko:2015yia}:
\begin{equation}
\label{eq:fug}
\lambda_q (\tau) = 1 - \exp\left(\frac{\tau_0-\tau}{{\tau_*}}\right)\,.
\end{equation}
Here $\tau_*$ characterizes the quark chemical
equilibration time.

To illustrate the above considerations, we apply the (2+1)-dimensional
relativistic hydrodynamics framework with a time-dependent equation of
state, developed in Refs.~\cite{Vovchenko:2016ijt,Vovchenko:2016mtf} and
implemented in the \texttt{vHLLE} package~\cite{Karpenko:2013wva}, to
$\text{p}\bar{\text{p}}$ collisions at HESR.  The equation of state interpolates linearly between the lattice equations of state for the
purely gluonic Yang-Mills (YM) theory~\cite{Borsanyi:2012ve}
$P_{\rm YM}(T)$ at $\lambda_q = 0$ and the full QCD with (2+1) quark flavors
\cite{Borsanyi:2013bia} $P_{\rm QCD}(T)$ at $\lambda_q = 1$:
\begin{align}
\label{eq:interpolated}
P(T, \lambda_q) & = \lambda_q \, P_{\rm QCD} (T) + (1 - \lambda_q) \, P_{\rm YM} (T) \nonumber \\
& = P_{\rm YM} (T) + \lambda_q \, [ P_{\rm QCD} (T) - P_{\rm YM} (T) ].
\end{align}
$P_{\rm YM}(T)$ and $P_{\rm QCD}(T)$ are shown in
Fig.~\ref{fig:latticeEoS}.

The hydrodynamic simulations of $\text{p} \bar{\text{p}}$ collisions at
$\sqrt{s} = 32\,\GeV$ discussed below assume a hard-sphere initial energy density
profile with radius $R = 0.6 \, \fm$.  The normalization of the energy
is fixed in order to yield an initial temperature of 273\,MeV in the
central cell, which is slightly above the critical temperature of 270\,
MeV.  This choice is motivated by Bjorken model based estimates at
$\sqrt{s} = 30 \,\GeV$ for small systems~\cite{VovchenkoThesis}.

\begin{figure}[!h]
\centering
\includegraphics[width=0.49\textwidth]{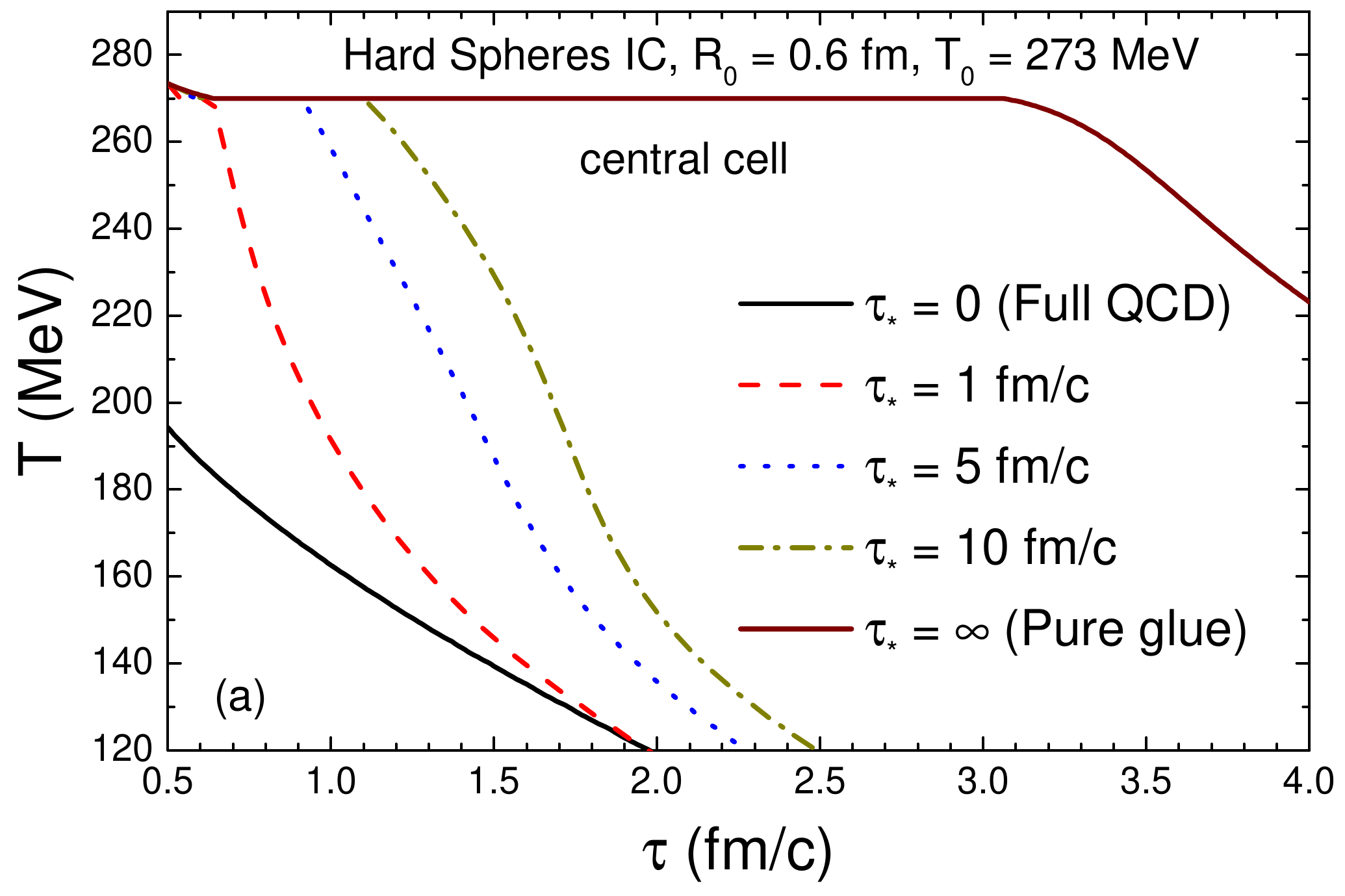}
\includegraphics[width=0.49\textwidth]{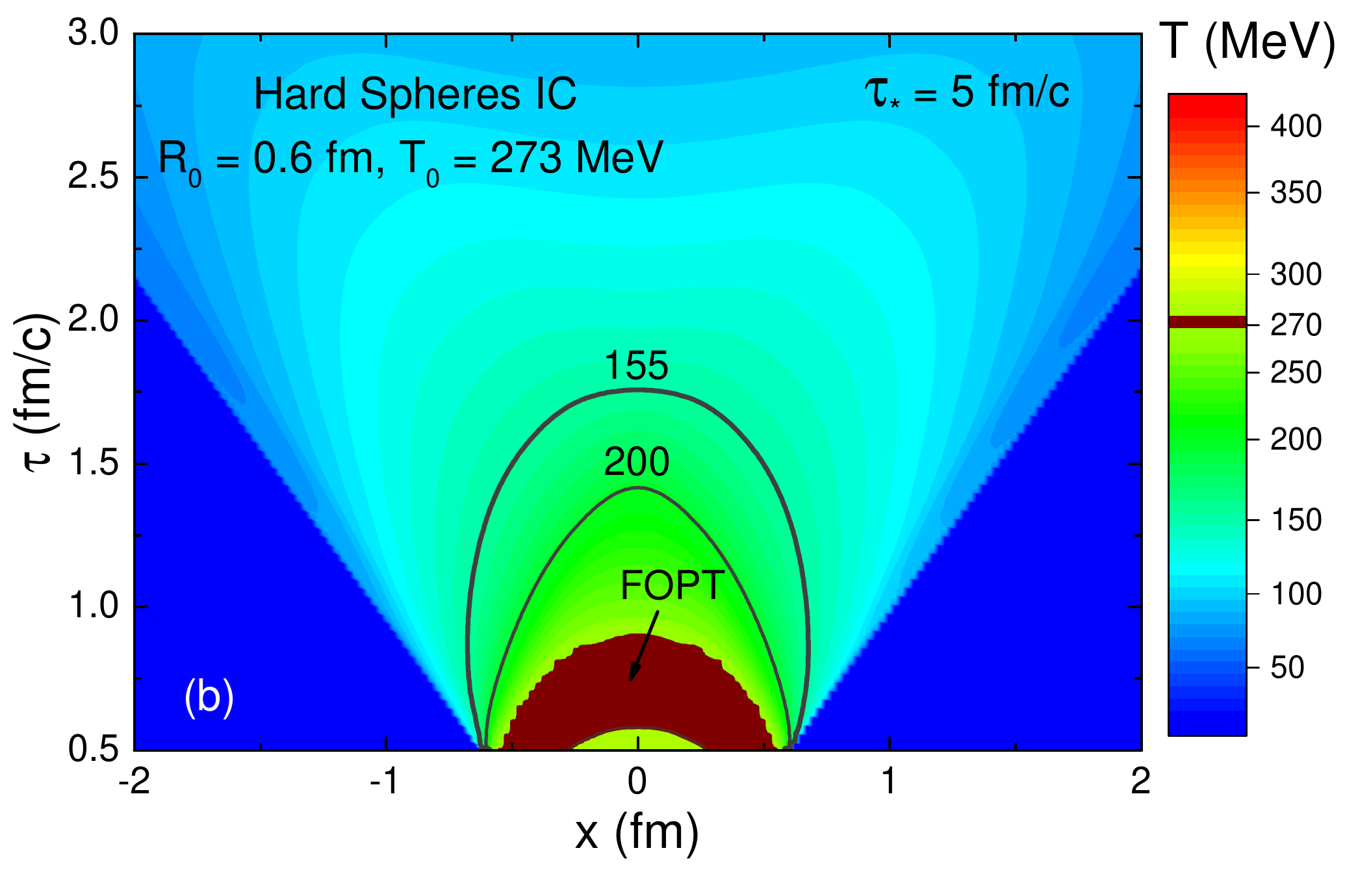}
\caption{The temperature profile of the central cell in the longitudinally boost invariant
  (2+1)-dimensional hydro evolution for $\text{p} \bar{\text{p}}$
  collisions in the pure glue initial state, the Yang-Mills scenario. 
  A hard spheres overlap, transverse density profile with radius $R = 0.6 \,\fm$ is used
  as the initial condition.  The normalization is fixed in order to
  yield the initial temperature of 273\,MeV in the central cell. (a) The
  $\tau$-dependence of the temperature is given for the central cell for different
  quark equilibration times: $\tau_* = 0$ (instant equilibration),
  $1\,\fm/c$ (fast equilibration), $5 \,\fm/c$ (moderate equilibration),
  $10 \,\fm/c$ (slow equilibration), and for $\tau_* \to \infty$ (pure
  gluodynamic evolution).  (b) The temperature profile in the $x-\tau$
  plane for $\tau_* = 5 \,\fm/c$.}
\label{fig:pp-hydro}
\end{figure}

Figure~\ref{fig:pp-hydro} (a) shows the $\tau$-dependence of the
temperature in the central cell for different quark equilibration times:
$\tau_* = 0$ (instant equilibration), $1 \,\fm/c$ (fast equilibration),
$5\,\fm/c$ (moderate equilibration), $10 \,\fm/c$ (slow equilibration),
and $\tau_* \to \infty$ (pure gluodynamic evolution). In the pure
gluodynamic scenario, $\tau_* \to \infty$, the system spends a very long
time in the mixed-phase region.  A fast quark equilibration shortens the
time period spent in the mixed phase significantly. Nevertheless, a
significant fraction of the system evolution takes place in the mixed
phase of the gluon-glueball deconfinement phase transition even at
presence of a moderately fast quark equilibration ($\tau_* = 5\,\fm/c$),
as illustrated by Fig.~\ref{fig:pp-hydro}b. Thus, significant effects of
the initial pure glue state on electromagnetic and hadronic observables
are expected for this collision setup.

These results illuminate the future HESR-collider option with the
central PANDA experiment detector as an exciting upgrade for FAIR, a promising option to search for even heavier 
glueballs and hadrons than envisioned for the fixed target mode, and for other new exotic states of matter.


\subsection{Low- and large mass dilepton production}

Low-mass lepton pair production has raised the interest in the field for decades.  
A quite robust theoretical
understanding \cite{Rapp:1999us,Rapp:2009yu,
  Endres:2015egk,Galatyuk:2015pkq,Staudenmaier:2017vtq,Linnyk:2015rco}
of dilepton production in heavy-ion collisions at various energies has been gained.
There dileptons play a special role as messengers from the early stages, as penetrating probes.

\begin{figure}[t]
\includegraphics[width=0.45\linewidth]{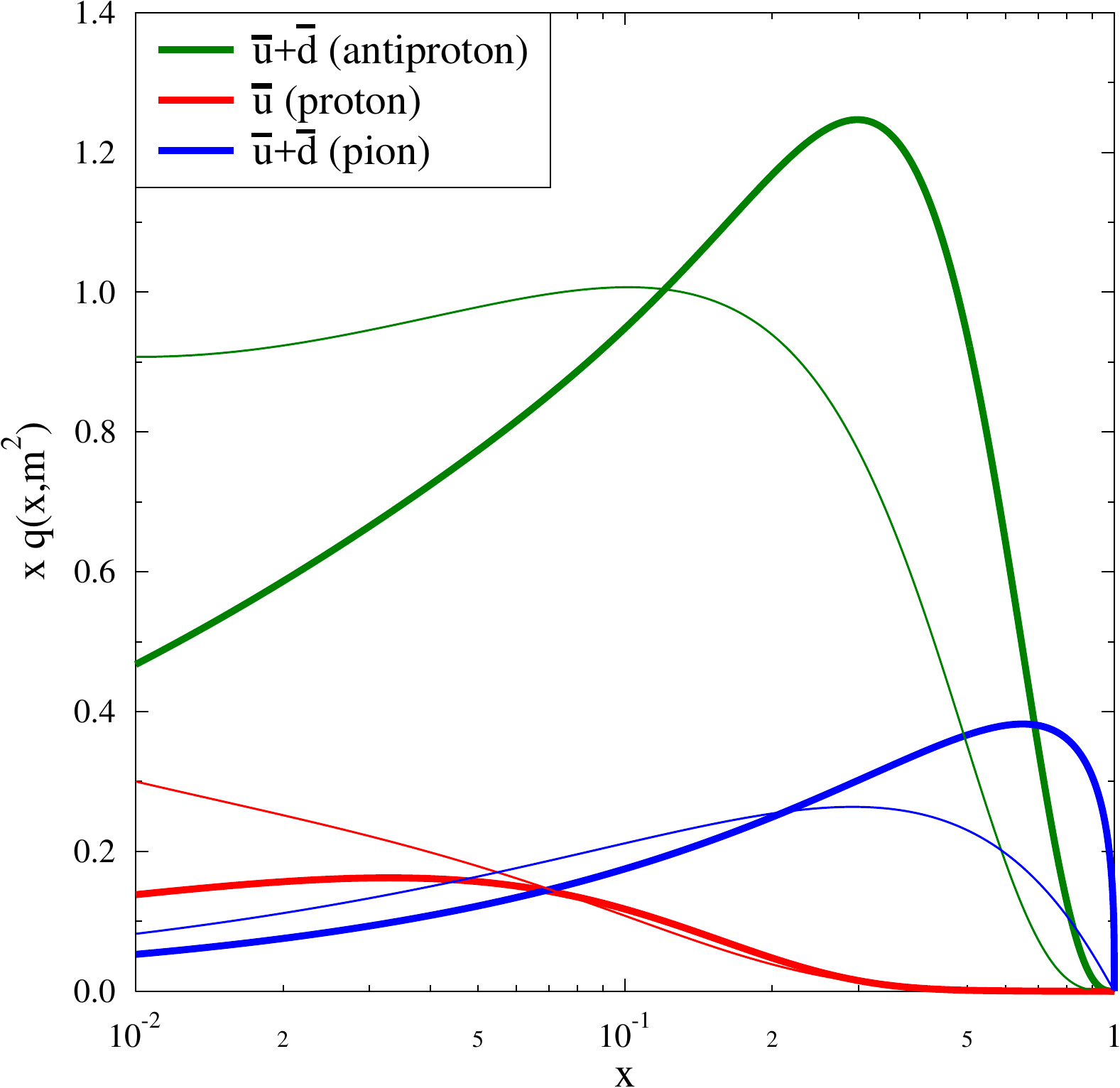} \hfill
\includegraphics[width=0.45\linewidth]{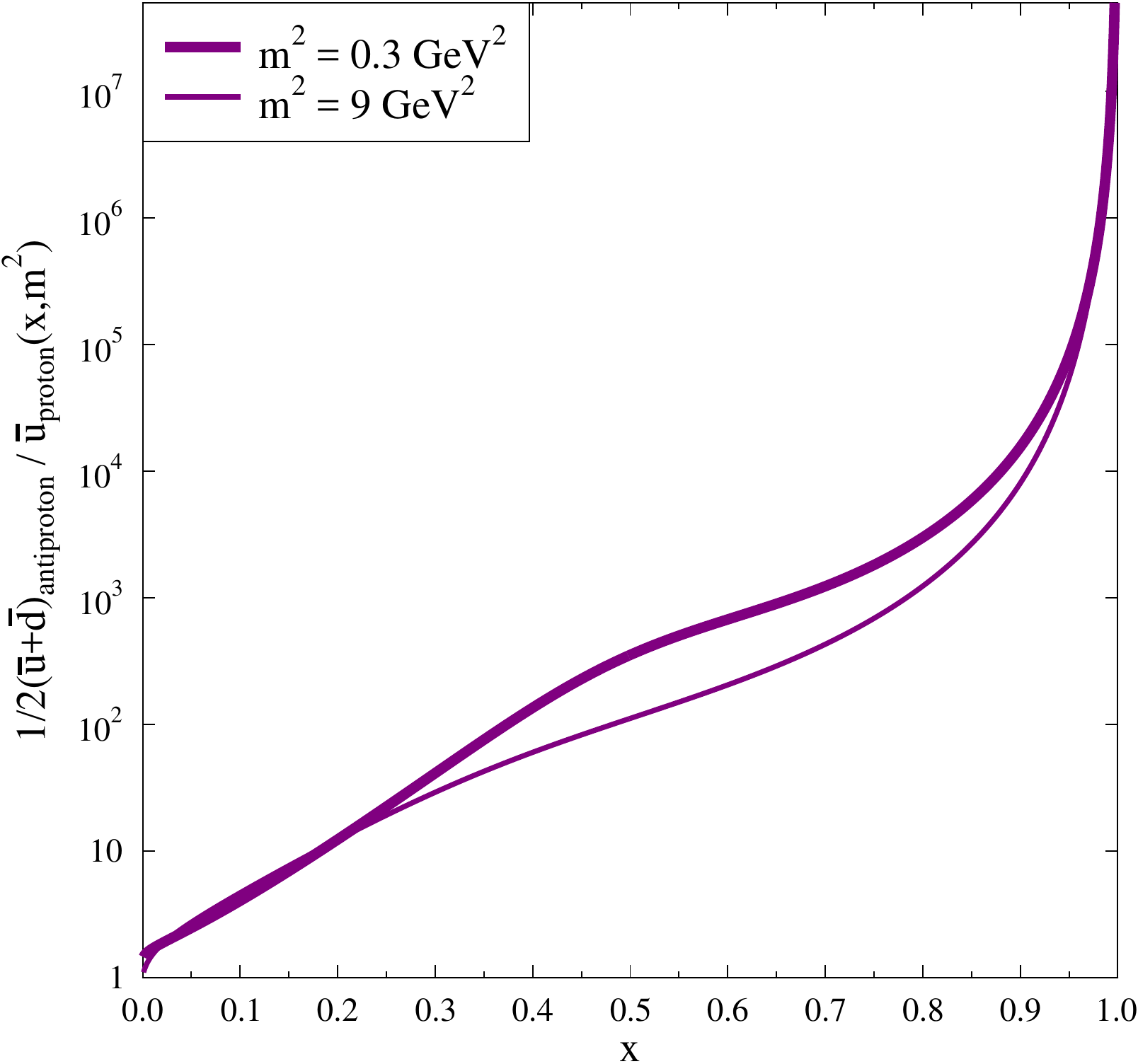}
\caption{Left: The light-anti-quark parton-distribution functions (PDFs)
  at $m^2=0.3 \;\GeV^2$ (thin lines) and $m^2=9 \; \GeV^2$ (bold lines)
  for anti-protons, protons \cite{Gluck:1994uf}, and pions
  \cite{Gluck:1991ey}. Right: The ratio of the PDFs for light
  anti-quarks in anti-protons and in protons.}
\label{fig.pdfs}
\end{figure}

At large invariant dilepton masses the perturbative  Drell-Yan (DY)
mechanism of QCD sets in. 
The minimal $M_{\text{DY}}^2$ value, where the DY mechanism dominates,
is not well known as  it is difficult to separate it experimentally
from the contribution of charm and $\text{J}/\uppsi$ production. 
It seems that the DY pair production mechanism dominates at $M_{\text{DY}}\geq 2\,\GeV$. 
For $\bar{\text{p}}$p(A) at intermediate
energies, $M_{\text{DY}}$ should be lower than for proton projectiles due to presence of abundant valence antiquarks.

Higher twist mechanisms may delay the inset of the leading-twist
contribution in the case of interactions with nuclei. 
Drell-Yan production of dileptons
at intermediate invariant masses by reactions with secondary mesons and
(anti-)baryons in p$A$ and $AA$ collisions has been investigated in \cite{Spieles:1997ih,Spieles:1998wz}. 
At the lower beam energies of the beam-energy scan at RHIC, at FAIR, and at NICA,
strong contributions from secondary DY processes are predicted for low-
and intermediate-mass dilepton pairs due to the formation of mesons at
time scales $\lesssim 1 \,\fm/c$, i.e. during the
interpenetration stage of projectile and target. 
This implies that
inflying primordial projectile and target nucleons from the
interpenetrating nuclei collide with just newly formed mesons (e.g., $\rho$
and $\omega$ with constituent-quark and constituent-antiquark masses of
$\sim 300 \, \MeV$ each).

For $\bar{\text{p}}\text{p}$ and $\bar{\text{p}} A$ reactions, the  Drell-Yan production is enhanced already for the
primordial collisions due to the
presence of \emph{valence} antiquarks of the antiproton, as displayed in Fig.\ \ref{fig.pdfs} which shows the ratio of light-antiquark PDFs in antiprotons and protons. 
With the HESR-C $\bar{\text{p}}$p and $\bar{\text{p}}A$ collider discussed here a direct assessment of the
valence-quark valence-antiquark parton distributions in
$\bar{\text{p}} \text{p}$ and $\bar{\text{p}} A$ collisions is accessible. 
The comparison of dilepton production by proton
and antiproton induced reactions provides unique information on the relative role of
initial- and final-state mechanisms. 
In particular, by comparing
dilepton production in proton and antiproton fragmentation
regions one may expect the maximal difference between the two cases.

Calculations of the inclusive (integrated over transverse momenta)
DY production can be performed now in  the NNLO  DGLAP
approximation. The techniques were developed for calculations of
transverse momentum distributions which include effects of multiple
gluon emissions (the Sudakov form factor effects) and
nonperturbative transverse momentum distributions (TMD), see e.g.  \cite{Angeles-Martinez:2015sea} and references therein.  TMDs
contribute to the differential cross section predominantly at small
transverse momenta of the DY pairs and for moderate masses of the pairs.

For large masses the theory works now very well providing a parameter free
description of the $Z-$boson production at the LHC including the
transverse momenta distribution.

\begin{figure}[t]
\includegraphics[angle=-90,keepaspectratio,width=0.44\linewidth]{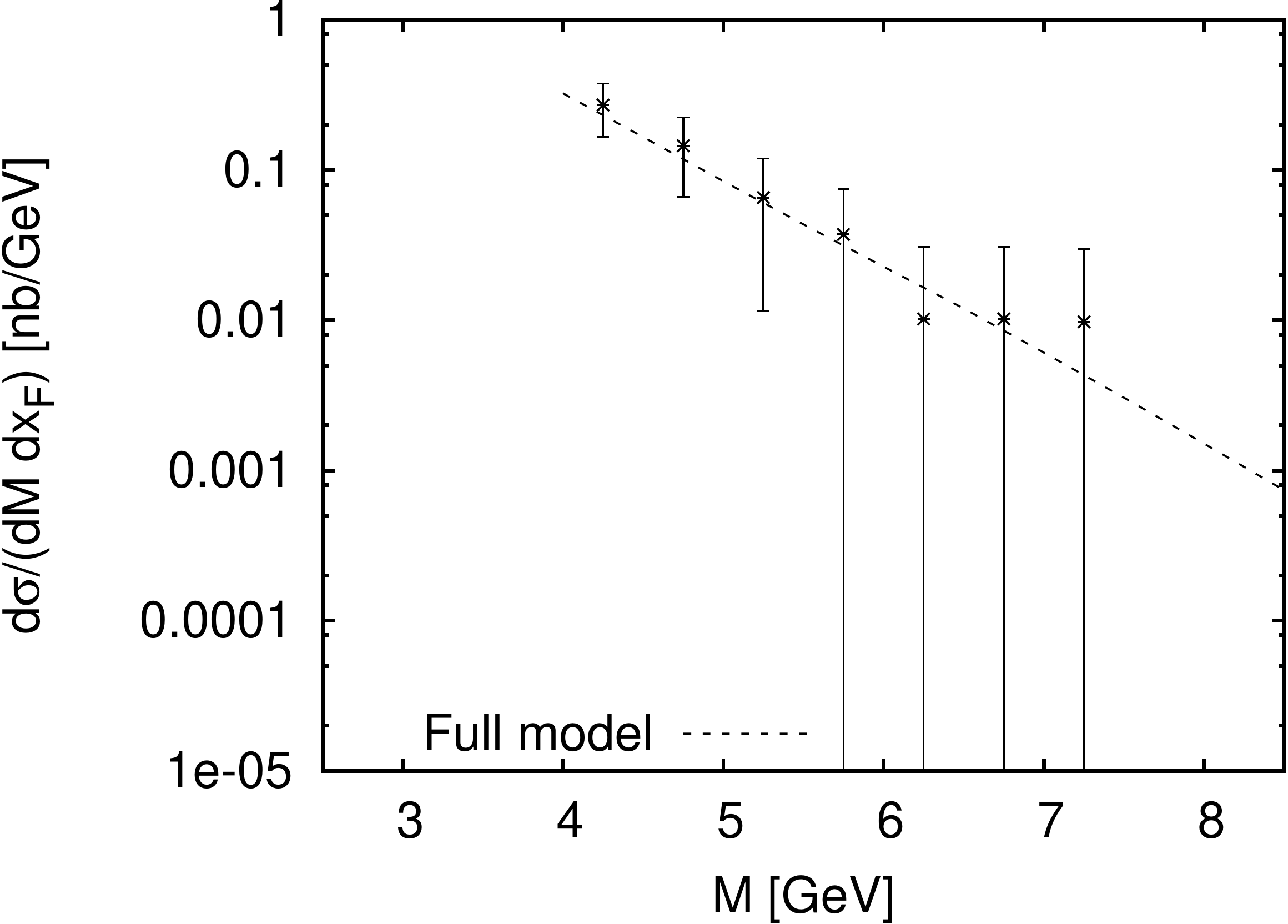} \hfill
  \includegraphics[angle=-90,keepaspectratio,width=0.44 \linewidth]{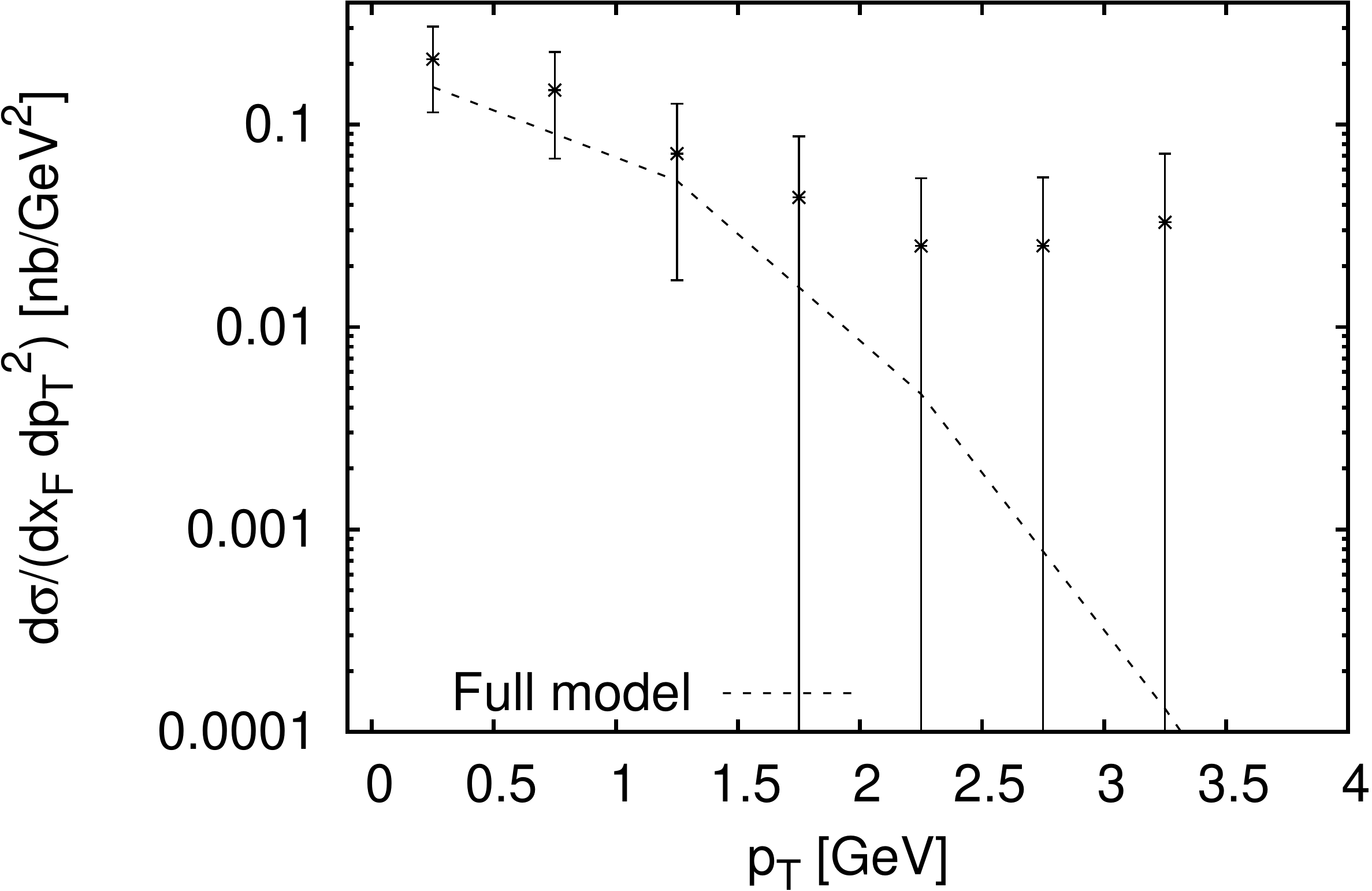}
    \caption{ Invariant-mass (left) and transverse-momentum
    spectra (right) of DY dimuon pairs in $\bar{\text{p}}\text{W}$ at
    $\sqrt{s}=15 \GeV^2$ collisions. Data are from the E537 collaboration
    \cite{Anassontzis:1987hk}. The used integrated PDF's are the
    MSTW2008LO68cl set \cite{Martin:2009iq}. 
     }
  \label{fig.pbar-W-E537-PANDA}
\end{figure}

Corresponding high precision data for fixed target energies are very
limited especially for the antiproton projectiles. Fig.\
\ref{fig.pbar-W-E537-PANDA} shows the result of the model for
$\bar{\text{p}}\text{W}$ collisions of the E537 collaboration at
$s=236 \GeV^2$ \cite{Anassontzis:1987hk}.

The data are compared with the model calculation
\cite{Eichstaedt:2011ke} which includes pQCD parton evolution, and the
TMD effects.

The model parameters have been fixed by using data on dimuon
transverse-momentum spectra in pp collisions at $s=1500 \, \GeV^2$ from
the E866 collaboration in \cite{Webb:2003ps,Webb:2003bj}. This model describes the dimuon production data without any
adjustment of model parameters quite well, in particular, the absolute cross section
in pd collisions from the E772 collaboration \cite{McGaughey:1994dx}, in
pCu collisions from the E605 collaboration \cite{Moreno:1990sf} at the
same collision energy, as well as in pA collisions from the E288
collaboration \cite{Ito:1980ev} and in pW collisions from the E439
collaboration \cite{Smith:1981gv} at $s=750~\GeV^2$.

The presence of the abundant antiquarks and the forward kinematics experimental arm of PANDA at the HESR-C
$\bar{\text{p}} \text{p}$ collider allows to determine the minimal $M^2$ for which the DY
mechanism works. (Note that the charm contribution is strongly
suppressed at $x_p \ge 0.4$ as a lepton in a charm decay carries, on
average, only 1/3 of the D-meson momentum (and even less for charm
baryons).

The onset of factorization in the DY process with nuclei  has still not
been explored, but will be very interesting. Indeed, as 
mentioned already, the formation length in these processes is pretty small. 
Hence, one could think of the process in a semiclassical way, for a
rather broad range of energies. In this picture, in the case of a nuclear
target, the antiproton may experience one or more (in)elastic
rescatterings on the target nucleons -- before it annihilates with a proton
into a dilepton pair. This effect can be taken into account within
transport models, e.g. GiBUU \cite{Buss:2011mx} calculations, which are a good setup for the antiproton-nucleus dynamics. The antiproton
stopping shall lead to a softening of the invariant-mass and
transverse-momentum spectra of the dilepton pairs. It is also important
to include these effects in future calculations, as it
influences the conclusions on the in-medium modifications of the nucleon PDFs.

At high enough energies the formation time becomes large, but an antiquark
which is involved in the DY process can experience energy losses growing quadratically with the path length~\cite{Baier:1996sk}.  
Here one expects $p_T$ broadening $\propto A^{1/3}$. 
These studies, if performed as a function of the atomic number and of the collision energy at fixed
$x_{\bar q}$,
may allow to explore these important effects in great detail.

\section{Final remarks}

The exciting science discussed here could be extended to much higher energies
and heavier states if -- in a later phase of FAIR -- 
one can manage to reinject antiprotons from the HESR into one or two of the higher energy main FAIR
synchrotrons, SIS 100 and SIS 300. Then $\bar{\text{p}}$A collisions can be synchronized to run
effectively and with high luminosity in a collider mode too. Also, crossing beams of SIS100 with SIS300, e.g. $45 \,A \GeV$
heavy ions in the SIS 300 colliding with $30 \,\GeV$ antiprotons (or
protons), or with ions of $15\, A\GeV$ in the SIS100 can be envisioned. Also other asymmetric
collisions maybe feasible, e.g., $90\,\GeV$ antiprotons/protons (SIS300)
colliding with $15 \, A\GeV$ heavy ions (or with $30\,\GeV$ protons or
antiprotons), in the SIS 100.

\section*{Acknowledgments}
We thank M.~Cacciari and R.~Vogt for discussions on charm and beauty production in $\bar{\text{p}} \text{p}$ scattering.
Research of L.F. and M.S. was supported by the US Department of Energy Office of Science, Office of Nuclear Physics under Award No. DE-FG02-93ER40771.
A.L. acknowledges partial financial support by Helmholtz International Center (HIC) for FAIR.
H.v.H. acknowledges the support from Frankfurt Institute for Advanced Studies~(FIAS).
H.St. acknowledges the support through the Judah M. Eisenberg Laureatus Chair by Goethe University  and the Walter Greiner Gesellschaft, Frankfurt.

\begin{flushleft}
\bibliographystyle{num-hvh-tit}
\bibliography{qftbib}

\begin{thebibliography}{10}
\providecommand{\url}[1]{\texttt{#1}}
\providecommand{\urlprefix}{}
\providecommand{\eprint}[2][]{\url{#2}}

\bibitem{Augustin:1974xw}
J.~E. Augustin et~al. (SLAC-SP-017 Collaboration), {Discovery of a Narrow
  Resonance in $\mathrm{e}^+ \mathrm{e}^-$ Annihilation}, Phys. Rev. Lett.
  \textbf{33}, 1406 (1974),
  \urlprefix\url{http://dx.doi.org/10.1103/PhysRevLett.33.1406}.

\bibitem{Aubert:1974js}
J.~J. Aubert et~al. (E598 Collaboration), {Experimental Observation of a Heavy
  Particle J}, Phys. Rev. Lett. \textbf{33}, 1404 (1974),
  \urlprefix\url{http://dx.doi.org/10.1103/PhysRevLett.33.1404}.

\bibitem{Herb:1977ek}
S.~W. Herb et~al., {Observation of a Dimuon Resonance at 9.5 GeV in 400 GeV
  Proton-Nucleus Collisions}, Phys. Rev. Lett. \textbf{39}, 252 (1977),
  \urlprefix\url{http://dx.doi.org/10.1103/PhysRevLett.39.252}.

\bibitem{Stocker:2015cva}
H.~St{\"o}cker, T.~St{\"o}hlker, and C.~Sturm, {FAIR - Cosmic Matter in the
  Laboratory}, J. Phys. Conf. Ser. \textbf{623}, 012026 (2015),
  \urlprefix\url{http://dx.doi.org/10.1088/1742-6596/623/1/012026}.

\bibitem{Barone:2005pu}
V.~Barone et~al. (PAX Collaboration), {Antiproton-proton scattering experiments
  with polarization}  (2005), \eprint{arXiv: hep-ex/0505054},
  \urlprefix\url{http://arxiv.org/hep-ex/0505054}.

\bibitem{pax:2006.asy2}
{PAX Collaboration}, {Technical Proposal for Antiproton-Proton Scattering
  Experiments with Polarization}, Tech. rep., Forschungszentrum J{\"u}lich
  (2006),
  \urlprefix\url{http://collaborations.fz-juelich.de/ikp/pax/public_files/proposals/techproposal20060125.pdf}.

\bibitem{Lehrbach:2007.asy3}
A.~Lehrach, {Accelerator Configuration for Polarized Proton-Antiproton Physics
  at FAIR}, AIP Conference Proceedings \textbf{915}, 147 (2007).

\bibitem{Mishustin:1993qg}
I.~N. Mishustin, L.~M. Satarov, J.~Schaffner, H.~St{\"o}cker, and W.~Greiner,
  {Baryon anti-baryon pair production in strong meson fields}, J. Phys. G
  \textbf{19}, 1303 (1993),
  \urlprefix\url{http://dx.doi.org/10.1088/0954-3899/19/9/009}.

\bibitem{Larionov:2008wy}
A.~B. Larionov, I.~N. Mishustin, L.~M. Satarov, and W.~Greiner, {Dynamical
  simulation of bound antiproton-nuclear systems and observable signals of cold
  nuclear compression}, Phys. Rev. C \textbf{78}, 014604 (2008),
  \urlprefix\url{http://dx.doi.org/10.1103/PhysRevC.78.014604}.

\bibitem{Bradamante:2005wk}
F.~Bradamante, I.~Koop, A.~Otboev, V.~Parkhomchuk, V.~Reva, P.~Shatunov, and
  Y.~Shatunov, {Conceptual design for a polarized proton-antiproton collider
  facility at GSI}  (2005), \eprint{arXiv: physics/0511252},
  \urlprefix\url{http://arxiv.org/abs/physics/0511252}.

\bibitem{Lehrach:2005ji}
A.~Lehrach, O.~Boine-Frankenheim, F.~Hinterberger, R.~Maier, and D.~Prasuhn,
  {Beam performance and luminosity limitations in the high-energy storage ring
  (HESR)}, Nucl. Instrum. Meth. A \textbf{561}, 289 (2006),
  \urlprefix\url{http://dx.doi.org/10.1016/j.nima.2006.01.017}.

\bibitem{Beller:2006sx}
P.~Beller, K.~Beckert, C.~Dimopoulou, A.~Dolinsky, F.~Nolden, M.~Steck, and
  J.~Yang, {Layout of an accumulator and decelerator ring for FAIR}, Conf.
  Proc. C \textbf{060626}, 199 (2006),
  \urlprefix\url{http://accelconf.web.cern.ch/AccelConf/e06/PAPERS/MOPCH074.PDF}.

\bibitem{Parkhomchuk:2004jc}
V.~V. Parkhomchuk, V.~B. Reva, A.~N. Skrinsky, V.~A. Vostrikov, K.~Beckert,
  P.~Beller, A.~Dolinskii, B.~Franzke, F.~Nolden, and M.~Steck, {An Electron
  Cooling System for the Proposed HESR Antiproton Storage Ring}, in \emph{{9th
  European Particle Accelerator Conference (EPAC 2004) Lucerne, Switzerland,
  July 5-9, 2004}} (2004),
  \urlprefix\url{http://accelconf.web.cern.ch/AccelConf/e04/PAPERS/WEPLT056.PDF}.

\bibitem{Reistad:2006vm}
D.~Reistad et~al., {Status of the HESR electron cooler design work}, Conf.
  Proc. C \textbf{060626}, 1648 (2006).

\bibitem{Kamerdzhiev:2014yza}
V.~Kamerdzhiev et~al., {2 MeV Electron Cooler for COSY and HESR -- First
  Results}, in \emph{{Proceedings, 5th International Particle Accelerator
  Conference (IPAC 2014): Dresden, Germany, June 15-20, 2014}}, MOPRI070
  (2014), \urlprefix\url{http://jacow.org/IPAC2014/papers/mopri070.pdf}.

\bibitem{Karliner:2015afa}
M.~Karliner, {Heavy exotic quarkonia and doubly heavy baryons}, EPJ Web Conf.
  \textbf{96}, 01019 (2015).

\bibitem{Cacciari:2005uk}
M.~Cacciari, P.~Nason, and C.~Oleari, {A Study of heavy flavored meson
  fragmentation functions in $\mathrm{e}^+ \mathrm{e}^-$ annihilation}, JHEP
  \textbf{04}, 006 (2006),
  \urlprefix\url{http://dx.doi.org/10.1088/1126-6708/2006/04/006}.

\bibitem{Beneke:2002ph}
M.~Beneke, A.~P. Chapovsky, M.~Diehl, and T.~Feldmann, {Soft collinear
  effective theory and heavy to light currents beyond leading power}, Nucl.
  Phys. B \textbf{643}, 431 (2002), \eprint{arXiv: hep-ph/0206152},
  \urlprefix\url{http://dx.doi.org.10.1016/S0550-3213(02)00687-9}.

\bibitem{CacciariVogt}
M. Cacciari and R. Vogt, private communications.

\bibitem{Tanabashi2018}
M.~Tanabashi et~al. (Particle Data Group), {The Review of Particle Physics
  (2018)}, Phys. Rev. D \textbf{98}, 030001 (2018),
  \urlprefix\url{http://pdg.lbl.gov/}.

\bibitem{Alkhazov:1978et}
G.~D. Alkhazov, S.~L. Belostotsky, and A.~A. Vorobev, {Scattering of 1-GeV
  Protons on Nuclei}, Phys. Rept. \textbf{42}, 89 (1978),
  \urlprefix\url{http://dx.doi.org/10.1016/0370-1573(78)90083-2}.

\bibitem{Larionov:2016xeb}
A.~B. Larionov and H.~Lenske, {Elastic scattering, polarization and absorption
  of relativistic antiprotons on nuclei}, Nucl. Phys. A \textbf{957}, 450
  (2017), \urlprefix\url{http://dx.doi.org/10.1016/j.nuclphysa.2016.10.006}.

\bibitem{Moskalenko:2001ya}
I.~V. Moskalenko, A.~W. Strong, J.~F. Ormes, and M.~S. Potgieter, {Secondary
  anti-protons and propagation of cosmic rays in the galaxy and heliosphere},
  Astrophys. J. \textbf{565}, 280 (2002),
  \urlprefix\url{http://dx.doi.org/10.1086/324402}.

\bibitem{Tyapkin:1976}
A.~A. Tyapkin, {A possible way of establishing the existence of charmed
  particles}, Sov. J. Nucl. Phys. \textbf{22}, 89 (1976).

\bibitem{Dover:1977jw}
C.~B. Dover and S.~H. Kahana, {Possibility of Charmed Hypernuclei}, Phys. Rev.
  Lett. \textbf{39}, 1506 (1977),
  \urlprefix\url{http://dx.doi.org/10.1103/PhysRevLett.39.1506}.

\bibitem{Tsushima:2002ua}
K.~Tsushima and F.~C. Khanna, {Lambda(c)+ and Lambda(b) hypernuclei}, Phys.
  Rev. C \textbf{67}, 015211 (2003),
  \urlprefix\url{http://dx.doi.org/10.1103/PhysRevC.67.015211}.

\bibitem{Larionov:2017hcm}
A.~B. Larionov and H.~Lenske, {Distillation of scalar exchange by coherent
  hypernucleus production in antiproton-nucleus collisions}, Phys. Lett. B
  \textbf{773}, 470 (2017),
  \urlprefix\url{http://dx.doi.org/10.1016/j.physletb.2017.09.007}.

\bibitem{Shyam:2016uxa}
R.~Shyam and K.~Tsushima, {Production of $\Lambda_c^+$ hypernuclei in
  antiproton - nucleus collisions}, Phys. Lett. B \textbf{770}, 236 (2017),
  \urlprefix\url{http://dx.doi.org/10.1016/j.physletb.2017.04.057}.

\bibitem{Gerland:1998bz}
L.~Gerland, L.~Frankfurt, M.~Strikman, H.~St{\"o}cker, and W.~Greiner,
  {J/$\psi$ production, $\chi$ polarization and color fluctuations}, Phys. Rev.
  Lett. \textbf{81}, 762 (1998), \eprint{arXiv:nucl-th/9803034},
  \urlprefix\url{http://dx.doi.org/10.1103/PhysRevLett.81.762}.

\bibitem{Frankfurt:1985cv}
L.~L. Frankfurt and M.~I. Strikman, {Point-like configurations in hadrons and
  nuclei and deep inelastic reactions with leptons: EMC and EMC-like effects},
  Nucl. Phys. B \textbf{250}, 143 (1985),
  \urlprefix\url{http://dx.doi.org/10.1016/0550-3213(85)90477-8}.

\bibitem{Alvioli:2014eda}
M.~Alvioli, B.~A. Cole, L.~Frankfurt, D.~V. Perepelitsa, and M.~Strikman,
  {Evidence for $x$-dependent proton color fluctuations in pA collisions at the
  CERN Large Hadron Collider}, Phys. Rev. C \textbf{93}, 011902 (2016),
  \urlprefix\url{http://dx.doi.org/10.1103/PhysRevC.93.011902}.

\bibitem{Alvioli:2017wou}
M.~Alvioli, L.~Frankfurt, D.~Perepelitsa, and M.~Strikman, {Global analysis of
  color fluctuation effects in proton- and deuteron-nucleus collisions at RHIC
  and the LHC}  (2017), \urlprefix\url{http://arXiv.org/abs/1709.04993}.

\bibitem{VanHove:1974wa}
L.~{Van Hove} and S.~Pokorski, {High-Energy Hadron-Hadron Collisions and
  Internal Hadron Structure}, Nucl. Phys. B \textbf{86}, 243 (1975),
  \urlprefix\url{http://dx.doi.org/10.1016/0550-3213(75)90443-5}.

\bibitem{Raha:1990dn}
S.~Raha, {Dilepton, diphoton and photon production in preequilibrium}, Phys.
  Scripta \textbf{T32}, 180 (1990),
  \urlprefix\url{http://dx.doi.org/10.1088/0031-8949/1990/T32/030}.

\bibitem{Shuryak:1992wc}
E.~V. Shuryak, {Two stage equilibration in high-energy heavy ion collisions},
  Phys. Rev. Lett. \textbf{68}, 3270 (1992),
  \urlprefix\url{http://dx.doi.org/10.1103/PhysRevLett.68.3270}.

\bibitem{Alam:1994sc}
J.~Alam, B.~Sinha, and S.~Raha, {Successive equilibration in quark - gluon
  plasma}, Phys. Rev. Lett. \textbf{73}, 1895 (1994).

\bibitem{Biro:1993qt}
T.~S. Biro, E.~van Doorn, B.~Muller, M.~H. Thoma, and X.~N. Wang, {Parton
  equilibration in relativistic heavy ion collisions}, Phys. Rev. C
  \textbf{48}, 1275 (1993),
  \urlprefix\url{http://dx.doi.org/10.1103/PhysRevC.48.1275}.

\bibitem{Elliott:1999uz}
D.~M. Elliott and D.~H. Rischke, {Chemical equilibration of quarks and gluons
  at RHIC and LHC energies}, Nucl. Phys. A \textbf{671}, 583 (2000),
  \urlprefix\url{http://dx.doi.org/10.1016/S0375-9474(99)00840-4}.

\bibitem{Xu:2004mz}
Z.~Xu and C.~Greiner, {Thermalization of gluons in ultrarelativistic heavy ion
  collisions by including three-body interactions in a parton cascade}, Phys.
  Rev. C \textbf{71}, 064901 (2005),
  \urlprefix\url{http://dx.doi.org/10.1103/PhysRevC.71.064901}.

\bibitem{Stoecker:2015zea}
H.~St{\"o}cker et~al., {Glueballs amass at RHIC and LHC Colliders! - The early
  quarkless 1st order phase transition at $T=270$ MeV - from pure Yang-Mills
  glue plasma to GlueBall-Hagedorn states}, J. Phys. G \textbf{43}, 015105
  (2016), \urlprefix\url{http://dx.doi.org/10.1088/0954-3899/43/1/015105}.

\bibitem{Stocker:2015nka}
H.~St{\"o}cker et~al., {Under‐saturation of quarks at early stages of
  relativistic nuclear collisions: The hot glue initial scenario and its
  observable signatures}, Astron. Nachr. \textbf{336} (2015),
  \urlprefix\url{http://dx.doi.org/10.1002/asna.201512252}.

\bibitem{Borsanyi:2012ve}
S.~Borsanyi, G.~Endrodi, Z.~Fodor, S.~D. Katz, and K.~K. Szabo, {Precision
  SU(3) lattice thermodynamics for a large temperature range}, JHEP
  \textbf{07}, 056 (2012), \eprint{1204.6184},
  \urlprefix\url{http://dx.doi.org/10.1007/JHEP07(2012)056}.

\bibitem{Beitel:2016ghw}
M.~Beitel, C.~Greiner, and H.~St{\"o}cker, {Fast dynamical evolution of a
  hadron resonance gas via Hagedorn states}, Phys. Rev. C \textbf{94}, 021902
  (2016), \urlprefix\url{http://dx.doi.org/10.1103/PhysRevC.94.021902}.

\bibitem{Vovchenko:2015yia}
V.~Vovchenko, M.~I. Gorenstein, L.~M. Satarov, I.~N. Mishustin, L.~P. Csernai,
  I.~Kisel, and H.~St{\"o}cker, {Entropy production in chemically
  nonequilibrium quark-gluon plasma created in central Pb+Pb collisions at
  energies available at the CERN Large Hadron Collider}, Phys. Rev. C
  \textbf{93}, 014906 (2016),
  \urlprefix\url{http://dx.doi.org/10.1103/PhysRevC.93.014906}.

\bibitem{Vovchenko:2016ijt}
V.~Vovchenko, I.~A. Karpenko, M.~I. Gorenstein, L.~M. Satarov, I.~N. Mishustin,
  B.~K{\"a}mpfer, and H.~St{\"o}cker, {Electromagnetic probes of a pure-glue
  initial state in nucleus-nucleus collisions at energies available at the CERN
  Large Hadron Collider}, Phys. Rev. C \textbf{94}, 024906 (2016),
  \urlprefix\url{http://dx.doi.org/10.1103/PhysRevC.94.024906}.

\bibitem{Vovchenko:2016mtf}
V.~Vovchenko, L.-G. Pang, H.~Niemi, I.~A. Karpenko, M.~I. Gorenstein, L.~M.
  Satarov, I.~N. Mishustin, B.~K{\"a}mpfer, and H.~St{\"o}cker, {Hydrodynamic
  modeling of a pure-glue initial scenario in high-energy hadron and heavy-ion
  collisions}, PoS \textbf{BORMIO2016}, 039 (2016).

\bibitem{Karpenko:2013wva}
I.~Karpenko, P.~Huovinen, and M.~Bleicher, {A 3+1 dimensional viscous
  hydrodynamic code for relativistic heavy ion collisions}, Comput. Phys.
  Commun. \textbf{185}, 3016 (2014),
  \urlprefix\url{http://dx.doi.org/10.1016/j.cpc.2014.07.010}.

\bibitem{Borsanyi:2013bia}
S.~Borsanyi, Z.~Fodor, C.~Hoelbling, S.~D. Katz, S.~Krieg, and K.~K. Szabo,
  {Full result for the QCD equation of state with 2+1 flavors}, Phys. Lett. B
  \textbf{730}, 99 (2014),
  \urlprefix\url{http://dx.doi.org/10.1016/j.physletb.2014.01.007}.

\bibitem{VovchenkoThesis}
V.~Vovchenko, \emph{{Quantum statistical van der Waals equation and its QCD
  applications}}, Ph.D. thesis, Goethe University Frankfurt (2018).

\bibitem{Rapp:1999us}
R.~Rapp and J.~Wambach, {Low mass dileptons at the CERN SPS: Evidence for
  chiral restoration?}, Eur. Phys. J. A \textbf{6}, 415 (1999),
  \urlprefix\url{http://dx.doi.org/10.1007/s100500050364}.

\bibitem{Rapp:2009yu}
R.~Rapp, J.~Wambach, and H.~van Hees, {The Chiral Restoration Transition of QCD
  and Low Mass Dileptons}, Landolt-B{\"o}rnstein \textbf{23}, 134 (2010),
  \urlprefix\url{http://dx.doi.org/10.1007/978-3-642-01539-7\_6}.

\bibitem{Endres:2015egk}
S.~Endres, H.~van Hees, and M.~Bleicher, {Photon and dilepton production at the
  Facility for Proton and Anti-Proton Research and beam-energy scan at the
  Relativistic Heavy-Ion Collider using coarse-grained microscopic transport
  simulations}, Phys. Rev. C \textbf{93}, 054901 (2016),
  \urlprefix\url{http://dx.doi.org/10.1103/PhysRevC.93.054901}.

\bibitem{Galatyuk:2015pkq}
T.~Galatyuk, P.~M. Hohler, R.~Rapp, F.~Seck, and J.~Stroth, {Thermal Dileptons
  from Coarse-Grained Transport as Fireball Probes at SIS Energies}, Eur. Phys.
  J. A \textbf{52}, 131 (2016),
  \urlprefix\url{http://dx.doi.org/10.1140/epja/i2016-16131-1}.

\bibitem{Staudenmaier:2017vtq}
J.~Staudenmaier, J.~Weil, V.~Steinberg, S.~Endres, and H.~Petersen, {Dilepton
  production and resonance properties within a new hadronic transport approach
  in the context of the GSI-HADES experimental data}  (2017), \eprint{arXiv:
  1711.10297 [nucl-th]}.

\bibitem{Linnyk:2015rco}
O.~Linnyk, E.~L. Bratkovskaya, and W.~Cassing, {Effective QCD and transport
  description of dilepton and photon production in heavy-ion collisions and
  elementary processes}, Prog. Part. Nucl. Phys. \textbf{87}, 50 (2016),
  \urlprefix\url{http://dx.doi.org/10.1016/j.ppnp.2015.12.003}.

\bibitem{Gluck:1994uf}
M.~Gl{\"u}ck, E.~Reya, and A.~Vogt, {Dynamical parton distributions of the
  proton and small x physics}, Z. Phys. C \textbf{67}, 433 (1995),
  \urlprefix\url{http://dx.doi.org/10.1007/BF01624586}.

\bibitem{Gluck:1991ey}
M.~Gluck, E.~Reya, and A.~Vogt, {Pionic parton distributions}, Z. Phys. C
  \textbf{53}, 651 (1992),
  \urlprefix\url{http://dx.doi.org/10.1007/BF01559743}.

\bibitem{Spieles:1997ih}
C.~Spieles, L.~Gerland, N.~Hammon, M.~Bleicher, S.~A. Bass, H.~St{\"o}cker,
  W.~Greiner, C.~Lourenco, and R.~Vogt, {A Microscopic calculation of secondary
  Drell-Yan production in heavy ion collisions}, Eur. Phys. J. C \textbf{5},
  349 (1998), \urlprefix\url{http://dx.doi.org/10.1007/s100520050279}.

\bibitem{Spieles:1998wz}
C.~Spieles, L.~Gerland, N.~Hammon, M.~Bleicher, S.~A. Bass, H.~St{\"o}cker,
  W.~Greiner, C.~Lourenco, and R.~Vogt, {Intermediate mass dileptons from
  secondary Drell-Yan processes}, Nucl. Phys. A \textbf{638}, 507 (1998),
  \urlprefix\url{http://dx.doi.org/10.1016/S0375-9474(98)00345-5}.

\bibitem{Angeles-Martinez:2015sea}
R.~Angeles-Martinez et~al., {Transverse Momentum Dependent (TMD) parton
  distribution functions: status and prospects}, Acta Phys. Polon. B
  \textbf{46}, 2501 (2015),
  \urlprefix\url{http://dx.doi.org/10.5506/APhysPolB.46.2501}.

\bibitem{Anassontzis:1987hk}
E.~Anassontzis et~al., {High mass dimuon production in $\bar{p} n$ and $\pi^-
  n$ interactions at 125-GeV/c}, Phys. Rev. D \textbf{38}, 1377 (1988),
  \urlprefix\url{http://dx.doi.org/10.1103/PhysRevD.38.1377}.

\bibitem{Eichstaedt:2011ke}
F.~Eichstaedt, S.~Leupold, K.~Gallmeister, H.~van Hees, and U.~Mosel,
  {Description of Fully Differential Drell-Yan Pair Production}, PoS
  \textbf{BORMIO2011}, 042 (2011), \eprint{arXiv: 1108.5287 [hep-ph]}.

\bibitem{Martin:2009iq}
A.~D. Martin, W.~J. Stirling, R.~S. Thorne, and G.~Watt, {Parton distributions
  for the LHC}, Eur. Phys. J. C \textbf{63}, 189 (2009),
  \urlprefix\url{http://dx.doi.org/10.1140/epjc/s10052-009-1072-5}.

\bibitem{Webb:2003ps}
J.~C. Webb et~al. (NuSea Collaboration), {Absolute Drell-Yan dimuon
  cross-sections in 800 GeV/$c$ pp and pd collisions}  (2003), \eprint{arXiv:
  hep-ex/0302019}.

\bibitem{Webb:2003bj}
J.~C. Webb, \emph{{Measurement of continuum dimuon production in 800 GeV/$c$
  proton nucleon collisions}}, Ph.D. thesis, New Mexico State U. (2003),
  \eprint{arXiv: hep-ex/0301031},
  \urlprefix\url{http://dx.doi.org/10.2172/1155678}.

\bibitem{McGaughey:1994dx}
P.~L. McGaughey et~al. (E772 Collaboration), {Cross-sections for the production
  of high mass muon pairs from 800-GeV proton bombardment of H-2}, Phys. Rev. D
  \textbf{50}, 3038 (1994), [Erratum: Phys. Rev.D60,119903(1999)],
  \urlprefix\url{http://dx.doi.org/10.1103/PhysRevD.50.3038,
  10.1103/PhysRevD.60.119903}.

\bibitem{Moreno:1990sf}
G.~Moreno et~al., {Dimuon production in proton - copper collisions at
  $\sqrt{s}$ = 38.8 GeV}, Phys. Rev. D \textbf{43}, 2815 (1991),
  \urlprefix\url{http://dx.doi.org/10.1103/PhysRevD.43.2815}.

\bibitem{Ito:1980ev}
A.~S. Ito et~al., {Measurement of the Continuum of Dimuons Produced in
  High-Energy Proton - Nucleus Collisions}, Phys. Rev. D \textbf{23}, 604
  (1981), \urlprefix\url{http://dx.doi.org/10.1103/PhysRevD.23.604}.

\bibitem{Smith:1981gv}
S.~R. Smith et~al., {Experimental Test of the {Drell-Yan} Model in $p W \to
  \mu^+ \mu^- X$}, Phys. Rev. Lett. \textbf{46}, 1607 (1981),
  \urlprefix\url{http://dx.doi.org/10.1103/PhysRevLett.46.1607}.

\bibitem{Buss:2011mx}
O.~Buss, T.~Gaitanos, K.~Gallmeister, H.~van Hees, M.~Kaskulov, et~al.,
  {Transport-theoretical Description of Nuclear Reactions}, Phys. Rept.
  \textbf{512}, 1 (2012),
  \urlprefix\url{http://dx.doi.org/10.1016/j.physrep.2011.12.001}.

\bibitem{Baier:1996sk}
R.~Baier, Y.~L. Dokshitzer, A.~H. Mueller, S.~Peigne, and D.~Schiff, {Radiative
  energy loss and p(T) broadening of high-energy partons in nuclei}, Nucl.
  Phys. B \textbf{484}, 265 (1997),
  \urlprefix\url{http://dx.doi.org/10.1016/S0550-3213(96)00581-0}.

\end{thebibliography}
\end{flushleft}

\end{document}